\documentclass[journal]{IEEEtran}
\usepackage{cite}
\usepackage{times}
\usepackage{epsfig}
\usepackage{graphicx}
\usepackage{amsmath}
\usepackage{amssymb}
\usepackage{epstopdf}
\usepackage{subfigure}
\usepackage{url}
\usepackage{multirow}
\usepackage{booktabs}
\usepackage{tabularx}
\usepackage[noend]{algpseudocode}
\usepackage{algorithmicx,algorithm}

\begin{document}

\title{Asymmetric Learned Image Compression with  Multi-Scale Residual Block, Importance Map, and Post-Quantization Filtering}
%
%
%
\author{Haisheng~Fu,
        Feng~Liang,
        Jie~Liang,
        Binglin~Li,
        Guohe~Zhang,
        Jingning Han
\thanks{Haisheng~Fu, Feng~Liang and Guohe~Zhang are with the School of Microelectronics, Xi'an Jiaotong University, Xi'an, China (e-mail: fhs4118005070@stu.xjtu.edu.cn; fengliang@xjtu.edu.cn;  zhangguohe@xjtu.edu.cn) (Corresponding author: Feng Liang)}
\thanks{Binglin Li and Jie Liang are with the School of Engineering Science, Simon Fraser University, Canada ( e-mail: binglinl@sfu.ca; jiel@sfu.ca)}
\thanks{Jingning~Han is with the Google Inc. (e-mail: jingning@google.com)}
\thanks{This work was supported by the National Natural Science Foundation of China (No. 61474093), the Natural Science Foundation of Shaanxi Province, China (No. 2020JM-006), the Natural Sciences and Engineering Research Council of Canada (RGPIN-2020-04525), China Scholarship Council, and Google Chrome University Research Program.} }

\markboth{Submitted to IEEE Transactions on Multimedia}%
{Shell \MakeLowercase{\textit{et al.}}: Bare Demo of IEEEtran.cls for IEEE Journals}
%

\maketitle

\begin{abstract}
Recently, deep learning-based image compression has made significant progresses, and has achieved better rate-distortion (R-D) performance than the latest traditional method, H.266/VVC, in both subjective metric and the more challenging objective metric. However, a major problem is that many leading learned schemes cannot maintain a good trade-off between performance and complexity. In this paper, we propose an efficient and effective image coding framework, which achieves similar R-D performance with lower complexity than the state of the art. First, we develop an improved multi-scale residual block (MSRB) that can expand the receptive field and is easier to obtain global information. It can further capture and reduce the spatial correlation of the latent representations. Second, a more advanced importance map network is introduced to adaptively allocate bits to different regions of the image. Third, we apply a 2D post-quantization filter (PQF) to reduce the quantization error, motivated by the Sample Adaptive Offset (SAO) filter in video coding. Moreover, we find that the complexity of encoder and decoder have different effects on image compression performance. Based on this observation, we design an asymmetric paradigm, in which the encoder employs three stages of MSRBs to improve the learning capacity, whereas the decoder only needs one stage of MSRB to yield satisfactory reconstruction, thereby reducing the decoding complexity without sacrificing performance. Experimental results show that compared to the state-of-the-art method, the encoding and decoding time of the proposed method are about 17 times faster, and the R-D performance is only reduced by less than 1$\%$ on both Kodak and Tecnick datasets, which is still better than H.266/VVC(4:4:4) and other recent learning-based methods. Our source code is publicly available at \url{https://github.com/fengyurenpingsheng}.
\end{abstract}


\IEEEpeerreviewmaketitle

\section{Introduction}

\IEEEPARstart{D}eep learning has been successfully applied to the field of image compression.  Similar to classical image coding, the learning-based framework also includes transform, quantization, and entropy coding, but each component is based on deep learning networks. These modules can be jointly optimized in an end-to-end manner. 

Most encoding/decoding networks are based on the autoencoder architecture \cite{end_to_end}, which is very suitable for the image compression application. To find an efficient compact representation of the input image, various approaches have been proposed, including the generalized divisive normalization (GDN) \cite{GDN}, residual blocks \cite{resblock, FU2021}, and non-local attention module \cite{NonLocal_attension, chen2021}. 
\begin{figure}[!t]
	\centering
		\includegraphics[scale=0.6]{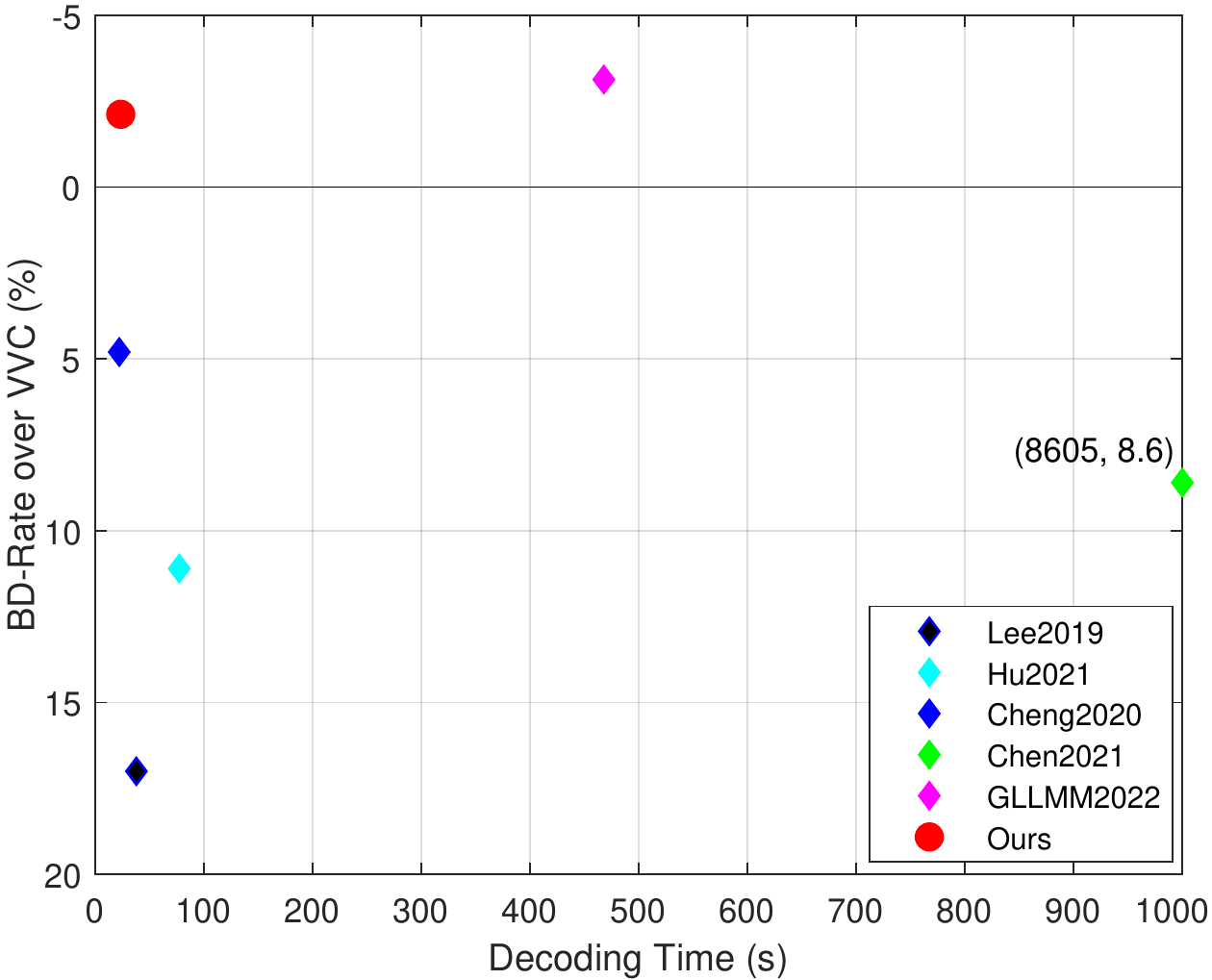}
	\caption{The BD-Rate and decoding time comparison of different methods using the Kodak dataset. The upper-left corner is better. The large decoding time of Chen2021 is written explicitly in the bracket. }
	\label{comp_time_bdrate}
\end{figure}

In the quantization step, a challenge for learned schemes is that the quantization operator is a non-differential process. To facilitate back propagation, different approximation methods have been proposed to make the quantification differentiable. Many works introduce additive uniform noise \cite{Variational, Joint, cheng2020} during training, and the rounding function is used for quantization during inference. In \cite{soft_to_hard, Fu_IEEE}, the soft-to-hard vector quantization is developed to replace scalar quantization.

In the entropy coding part, since it is difficult to calculate the actual entropy, the popular approach is to use context-adaptive entropy coding, which assumes that the latent representation obeys certain probability distribution, such as Gaussian mixture models \cite{Lossy_image, Variational, Joint}, Laplace mixture module \cite{tuya_2019}, Logistic mixture module \cite{pixelCNN++, L3C}, and Gaussian-Laplacian-Logistic mixture model (GLLMM) \cite{GLLMM}. The parameters of these models are estimated using a hyperprior network \cite{Variational}. In \cite{Joint}, the autoregressive mask convolution and the hyperprior are combined to further improve the performance. This approach has been adopted by most leading schemes.


Despite the promising results achieved in learned image coding, a major problem is that many leading learned schemes can not achieve a good trade off between complexity and R-D performance. The GLLMM-based scheme in \cite{GLLMM} has the state-of-the-art performance in learned image coding, which also outperforms the latest traditional approach, H.266/VVC intra coding (4:4:4 and 4:2:0), in both subjective metric (MS-SSIM) and the more challenging objective metric (PSNR). In addition to the GLLMM model, it also uses concatenated residual blocks (CRB) to improve the performance of the core encoding/decoding networks. However, the encoding and decoding time of GLLMM model is much longer than other models. Although some works \cite{cheng2020, Hu_2021, Lee_2021} have acceptable coding complexity, the compression performance is below H.266/VVC intra coding (4:4:4). Figure \ref{comp_time_bdrate} gives the rate-speed comparison among different methods.

In this paper, our goal is to propose an efficient and effective framework, which can achieve better compression performance with relatively low complexity. The technical contributions are summarized as follows.  

$\bullet\,$We develop a modified multi-scale residual block (MSRB) (\cite{MSRB_ECCV18}),  which can expand the receptive field and is easier to obtain global information. The modified MSRB can further capture and reduce spatial correlation of the latent representations and boost the compression performance.

$\bullet\,$We propose a more advanced importance map to adaptively allocate bits to different regions of the image. A channel rate allocation algorithm is provided, where the channels with non-zero coefficients will be encoded and decoded. The importance map module can produce more zeros, save more bits, and reduce the encoding and decoding time.

$\bullet\,$We apply a 2D post-quantization filter (PQF) after entropy decoding and before the decoding network to reduce the quantization error, motivated by the Sample Adaptive Offset (SAO) filtering in H.265/HEVC and H.266/VVC \cite{SAO_chuamin}. The difference is that our PQF is pre-trained and applied to the low-dimensional latent representation instead of the high-dimensional image domain.

$\bullet\,$We find that the complexity of the encoder and decoder have different effects on image compression performance. The image compression performance will not drop when we appropriately reduce the complexity of the decoder. This allows us to adopt an asymmetric design paradigm, in which the encoder employs three stages of MSRBs to improve the learning capacity, whereas the decoder only needs one stage of MSRB to yield satisfactory reconstruction, thereby reducing the decoding complexity. This paradigm is very useful for real-time image and video applications.


Experimental results using the Kodak and Tecnick datasets show that compared to the state-of-the-art GLLMM method in \cite{GLLMM}, the encoding and decoding of the proposed method are about 17 times faster, and the R-D performance is only reduced by less than 1.1$\%$ as shown in Figure \ref{comp_time_bdrate}. The proposed scheme still outperforms other state-of-the-art learning-based methods and H.266/VVC. Therefore it achieves the new state of the art in learned image coding when considering both the complexity and performance.

\begin{figure*}[!thp]
	\centering
		\includegraphics[scale=0.55]{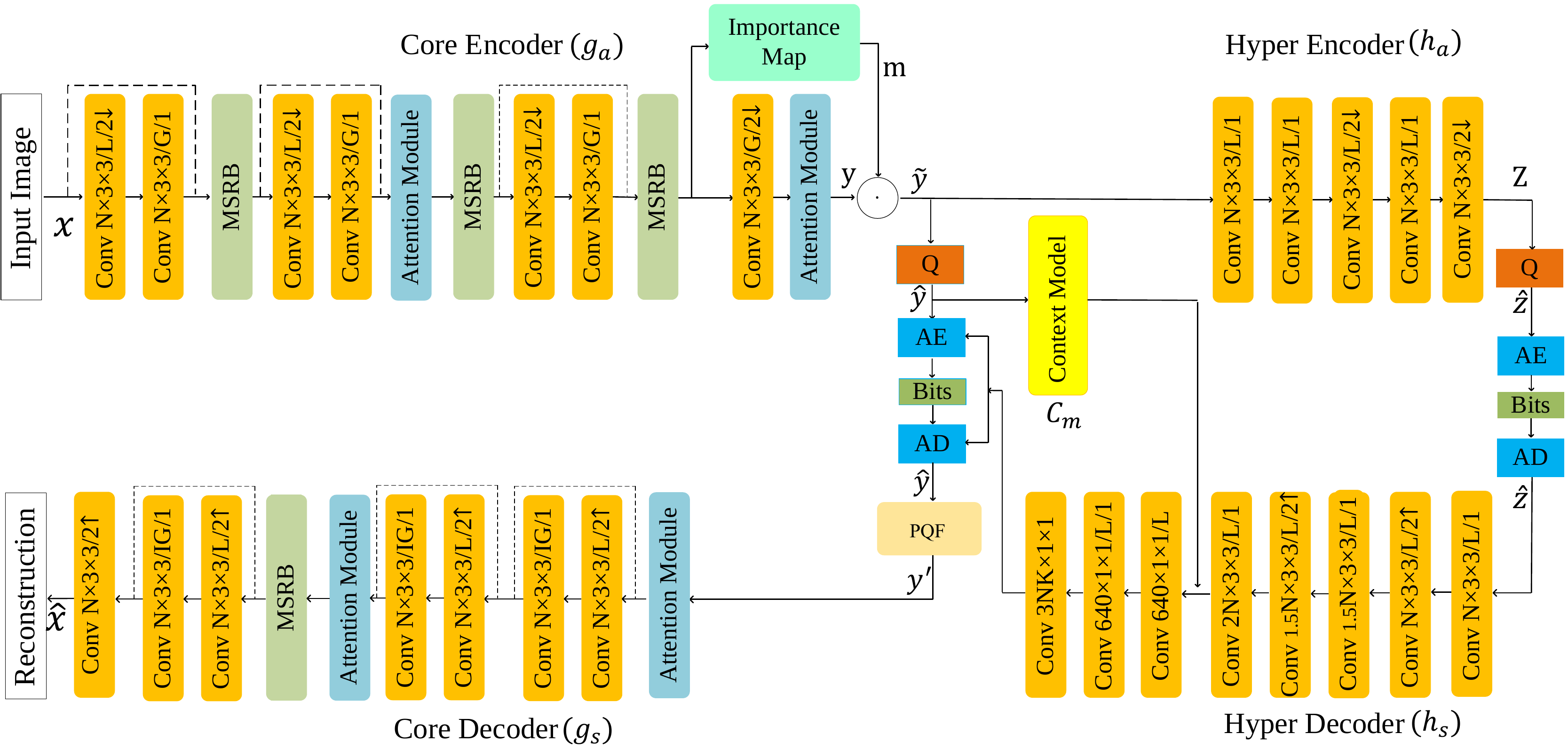}
	\caption{The detailed structures of the proposed image compression network. $G$ and $IG$ stands for GDN and inverse GDN, respectively. $\uparrow$ and $\downarrow$ represent the up- or down- sampling operation.  $3 \times 3$ represents the convolution kernel size. $AE$ and $AD$ respectively stand for arithmetic encoder and arithmetic decoder. $PQF$ denotes 2D Post-Quantization Filter. $L$ is leaky ReLU activation function. The dotted lines represent the shortcut connection with size change. $\bigodot$ represents the dot multiplication operation. }
	\label{networkstructure}
\end{figure*}

The remainder of the paper is organized as follows.  In Section  \ref{Related_work}, we briefly introduce some related works. In Section \ref{the_proposed_structure},  the proposed asymmetric learned image compression framework is proposed. In Section \ref{Implementation} and \ref{Experiment}, we give the implementation details and compare our methods with some traditional state-of-the-art learning-based methods and classical image compression methods. Ablation experiments are carried out to investigate the performance gain of the the proposed scheme. The conclusions are reported in Section \ref{Conclusion}.

\section{Related Work}
\label{Related_work}

\textbf{Autoencoder.} Most learned image compression methods are based on the autoencoder architecture. In \cite{GDN,end_to_end}, the generalized divisive normalization (GDN) is adopted. To enhance the performance of the network, the residual block first proposed in the ResNet is widely used in image compression \cite{resblock}, where a residual mapping with a shortcut is developed, which is easier to train than a direct mapping, and can enhance the learning ability of the network and reduce the spatial redundancy of latent representation. Some representative designs are \cite{NonLocal_attension, cheng2020, Content_Weighted, GLLMM}. In \cite{GLLMM}, a concatenated residual blocks (CRB) is developed. In \cite{MSRB_ECCV18}, a milti-scale residual block (MSRB) is developed for image super-resolution. In this paper, we develop a modified MSRB and apply it to image compression.

\textbf{Entropy Coding.} In \cite{Variational}, a hyperprior network is proposed to estimate the model parameters of the latent representation, which can facilitate the entropy coding. It achieves comparable performance to the H.265/HEVC-based BPG (4:4:4) codec in terms of the PSNR metric. In \cite{Joint}, an autoregressive mask convolution is further employed in order to get context-adaptive entropy coding. It is the first learned image coding scheme that can outperform BPG (4:4:4) in PSNR. In \cite{Lee_2020}, two types of contexts, bit-consuming contexts and bit-free contexts, are further developed. In \cite{cheng2020} and \cite{GLLMM}, the performance is further improved by introducing advanced entropy coding models, such as GMM and GLLMM. 


\textbf{Importance map.} Many images contain a lot of smooth background regions. We can allocate less bits to these areas without affecting the visual quality. The bits saved can be allocated to other complex and salient regions. To obtain the complexity and saliency information of an image, the spatially variant bit allocation methods are proposed. In \cite{Limu_conf,Content_Weighted}, the importance map network is developed to extract the importance map of the input image, but additional bits are needed to store the importance map. In \cite{Conditional}, the importance map is automatically optimized via the loss function.

\textbf{Quantization.} To make the quantization differentiable, in the training of many learned image coding schemes, an additive random uniform noise in the range of $[-0.5, 0.5]$ is added to the latent representation. During inference, the rounding function is employed to quantize the latent representation. However, this causes a mismatch between the training and inference, in addition to the quantization error itself. 

\textbf{SAO Filtering.} In H.265/HEVC and H.266/VVC, a new coding tool called Sample Adaptive Offset (SAO) filtering is added after decoding to further reduce the reconstruction error \cite{SAO_chuamin}. In \cite{SAO_fu, SAO_jia, SAO_Lee}, deep learning-based modules are developed to replace the traditional SAO filter, in order to improve the reconstruction performance.  

\section{The Proposed Asymmetric Learned Image Compression Framework}
\label{the_proposed_structure}

Fig. \ref{networkstructure} illustrates the proposed network architecture. The input image has a size of $W\times H\times 3$, where $W$ and $H$ are the length and width of the input image. The pixel values of the input image are scaled from $[0, 255]$ to $[-1, 1]$. Similar to \cite{Joint, cheng2020, GLLMM}, the architecture consists of two sub-networks: the core subnetworks and the hyper subnetworks. The core encoder network learns a quantized latent representation of the input image, while the entropy subnetwork learns the probabilistic model of the quantized latent representation, which is utilized in the entropy coding.

The core encoder network includes the backbone network and importance map subnetwork. The backbone network contains various convolution layers and four stages of pooling operators to obtain the low-dimensional latent representation $y$. The GDN operator is adopted when the size is changed. 

In the encoder backbone network, three stage of improved muiti-scale residual blocks (MSRBs) are used to improve the compression performance of the network, whose details will be explained in Sec. \ref{sec_msrb}. In \cite{GLLMM, cheng2020} and other existing schemes, the decoder network is usually symmetric to the encoder network, i.e., they have the same structure and same complexity. In fact, in many schemes, the decoder network is even more complicated than the encoder network, due to the use of some denoising networks to improve the reconstruction quality. However, our experimental results show that the symmetric structure can be simplified. We observe that the complexity of the encoder and decoder have different effects on image compression performance. The image compression performance will not drop when we appropriately  reduce the complexity of the decoder, which will be proved in Sec. \ref{Ablation}. Therefore, we propose an asymmetric design paradigm, where the decoder network is simpler than the encoder network. In this paper, instead of using three stages of MSRBs in the decoder network as in \cite{GLLMM, cheng2020}, we only use one stage of MSRB.  Experimental results in Sec. \ref{Experiment} will confirm that this asymmetric approach can still achieve the state-of-the-art performance, but with lower decoding complexity.

The asymmetric paradigm with lighter decoder than encoder is very attractive to practical applications, especially for video communications, because the decoder needs to decode the video in real time.

In the encoder and decoder, the simplified attention module in \cite{cheng2020} is also employed at two resolutions to enhance the learning ability of the network.

At the end of the encoder network, an importance map subnetwork is used to obtain the importance map $m$, which is element-wise multiplied to the latent representation $y$ from the encoder network. The masked result $\tilde{y}$ is then quantized and dequantized to get the reconstruction $\hat{y}$. Next, arithmetic coding is used to compress $\hat{y}$ into the bitstream. The decoded $\hat{y}$ is sent to the main decoder to generate the reconstructed image $x'$.

\begin{figure*}[tp]
\subfigure[]
{\begin{minipage}[t]{0.5\linewidth}
\centering
\includegraphics[scale=0.65]{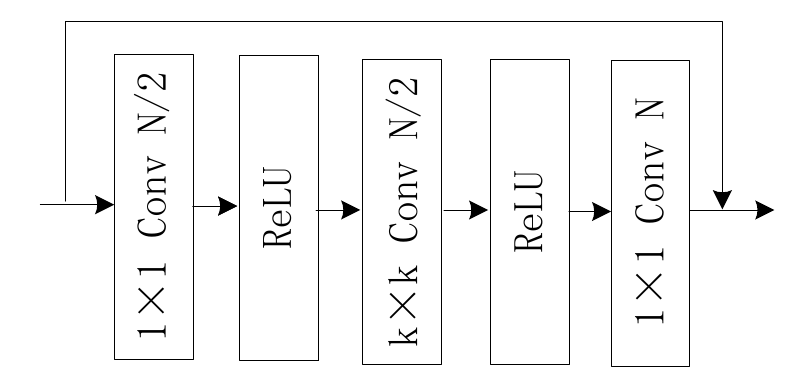}
\label{fig:residual_block}
\end{minipage}
}%
\subfigure[]
{\begin{minipage}[t]{0.5\linewidth}
\includegraphics[scale=0.65]{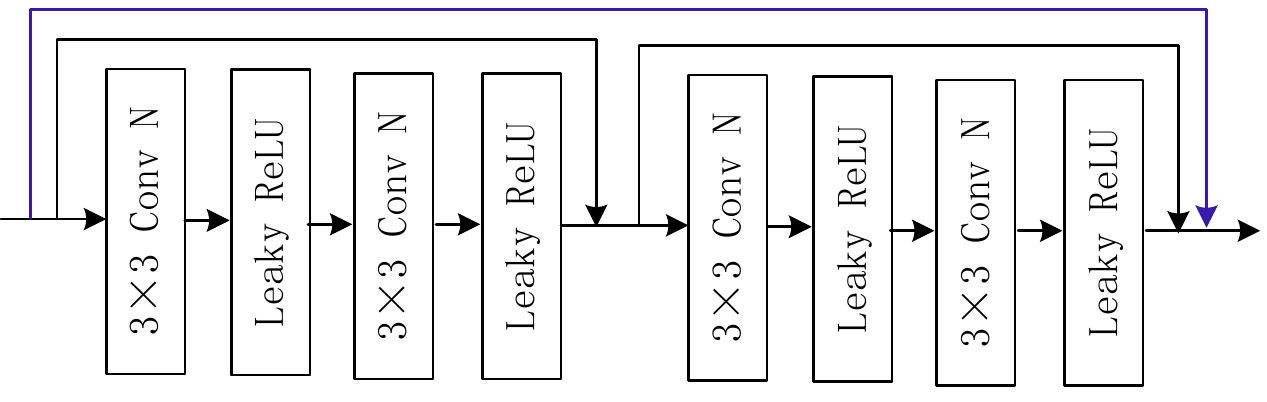}
\label{CRM}
\end{minipage}
}%

\centering
\subfigure[]
{\begin{minipage}[t]{0.5\linewidth}
\centering
\includegraphics[scale=0.65]{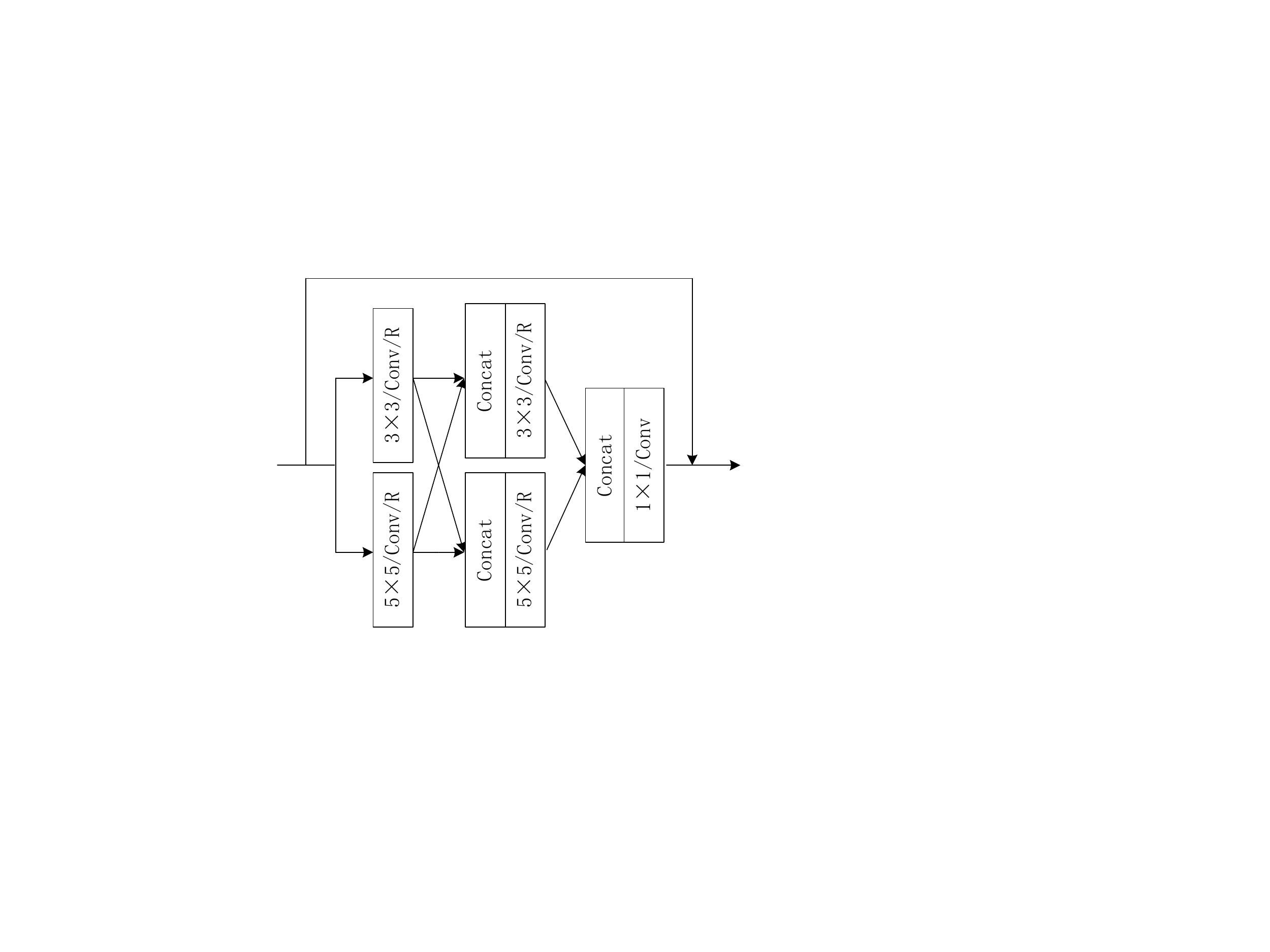}
\label{fig:msrb}
\end{minipage}
}%
\subfigure[]
{\begin{minipage}[t]{0.5\linewidth}
\includegraphics[scale=0.75]{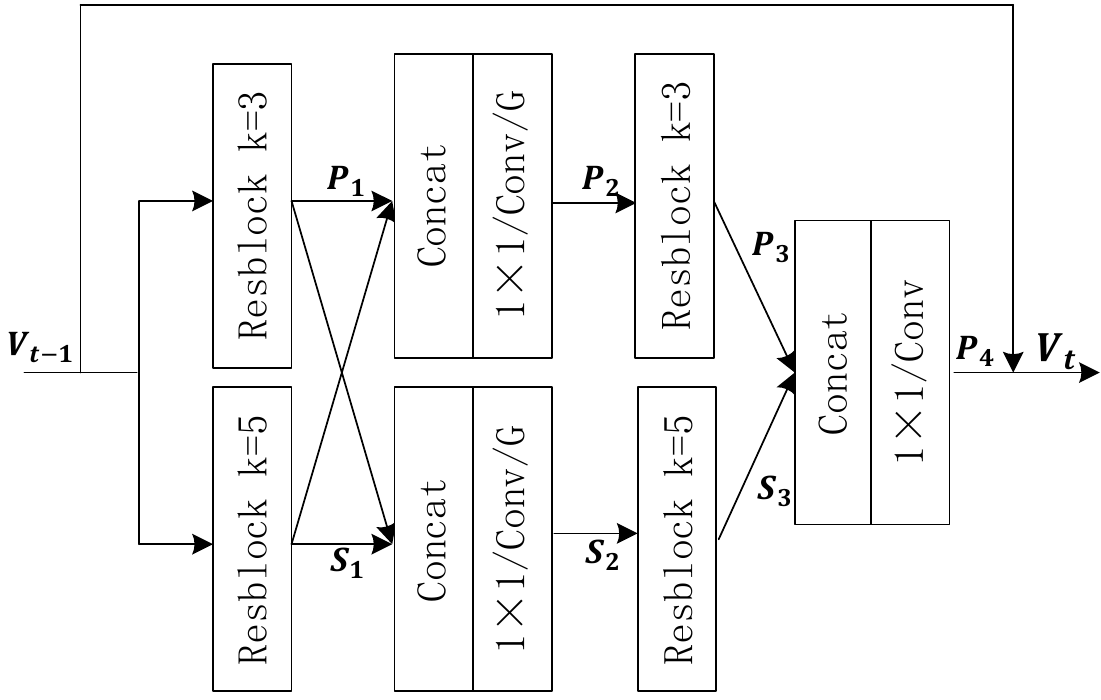}
\label{fig:multScale}
\end{minipage}
}%
\caption{(a) The residual block in \cite{resblock}. (b) The concatenated residual block (CRB) in \cite{GLLMM}. (c) The multi-scale residual block (MSRB) in \cite{MSRB_ECCV18}. (d) The proposed improved MSRB. $G$ represents GDN operator. $k$ is the convolution kernel size.}
\end{figure*}

To help the entropy coding, the hyper networks learn the distribution of the latent representation. In \cite{cheng2020}, the Gaussian mixture model (GMM) is used. In \cite{GLLMM}, a more accurate Gaussian-Laplacian-Logistic mixture model is developed. However, its complexity is much higher than the GMM model. In this paper, we adopt the simpler GMM model. Thanks to the following modifications, we can achieve comparable performance to \cite{GLLMM}, with much lower complexity. Our performance is also better than \cite{cheng2020}, with similar complexity.

The leaky ReLU is utilized in most convolution layers, except for the last layer in hyperprior encoder and decoder, which does not have any activation function.

\subsection{Multi-Scale Residual Block (MSRB)}
\label{sec_msrb}

The residual block proposed in \cite{resblock} has been widely used in many fields, including learned image compression. Its structure is show in Fig. \ref{fig:residual_block}. In \cite{GLLMM}, a concatenated residual blocks (CRB) is introduced, in which multiple residual blocks are serially connected with additional shortcut connections, as shown in Fig. \ref{CRM}.

In \cite{MSRB_ECCV18}, a multi-scale residual block (MSRB) was developed for image super-resolution, as shown in Fig. \ref{fig:msrb}. It was motivated by the inception module in \cite{goole_net}, where convolutions at different scales are used to expand the receptive field.

In this paper, we propose an improved MSRB and apply it to image compression, as shown in Fig. \ref{fig:multScale}. The MSRB in \cite{MSRB_ECCV18} only uses simple convolutions. In our MSRB, we replace it by the residual block in ResNet \cite{resblock}. The improved MSRB includes a two-branch network with different convolutional kernel size $k$, which can obtain global information at different scales. After the first residual block, the outputs of the two branches are concatenated, and processed separately, before being concatenated and processed again. A shortcut is also added between the input and the final output. In our MSRB, 1x1 filters are used extensively to control the complexity. The GDN operator is also used to improve the feature extraction.

As mentioned before, the MSRB is used asymmetrically in the encoder and the decoder. The encoder has three stages of MSRBs to ensure its learning capacity, whereas the decoder only needs one stage of MSRB, which can already achieve satisfactory reconstruction quality. This can reduce the complexity of the decoder, which is very important for many real-time image and video applications.


 \begin{figure}[!tp]
	\centering
		\includegraphics[scale=0.35]{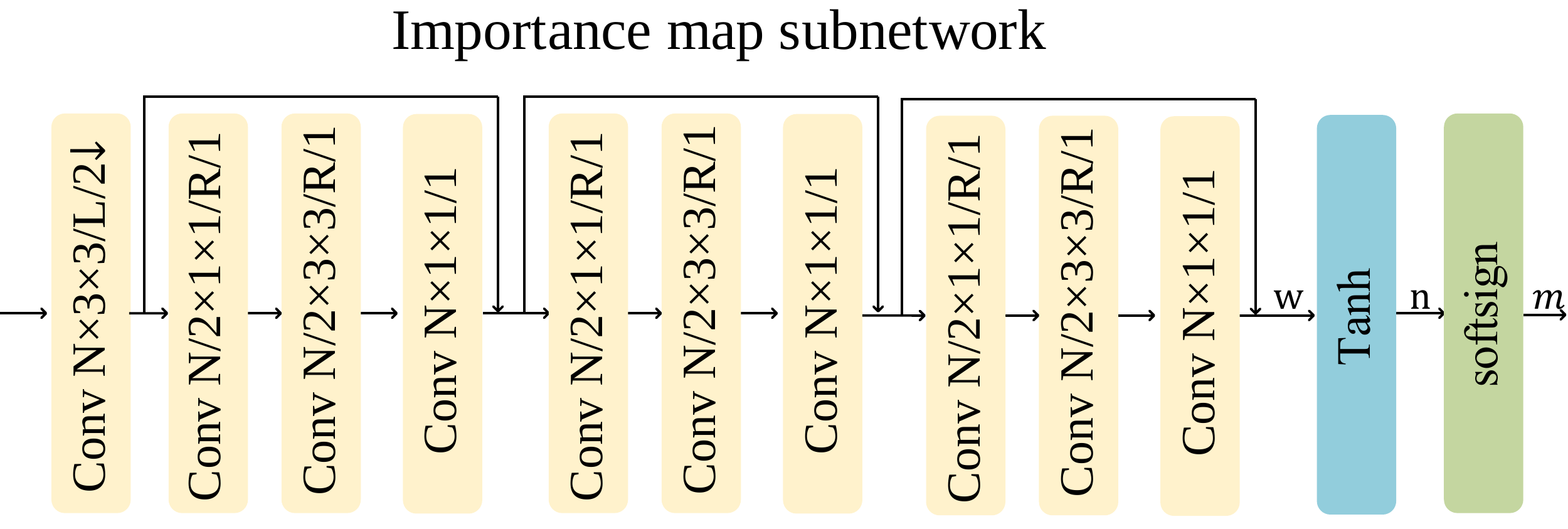}
	\caption{The detailed structures of the proposed importance map subnetwork. The symbol definitions are the same as in Fig. \ref{networkstructure}.}
	\label{BitAllocationSubnetworks}
\end{figure}

\subsection{Importance Map}
\label{importance map}

Images are generally composed of different contents. Therefore, it is desired to assign different bit rates in different regions.

In \cite{Conditional}, an importance map is developed to control the bit allocation in different regions, where the first channel of the latent representation is considered as the importance map. To control the bit allocations in different channels, it is mapped to other channels via Eq. \ref{mask}.
 \begin{equation}\label{mask}
  m_{i,j,c} = \left \{\begin{array}{ccl}
  1 &   &{if \quad c < y_{i,j}}\\
  (y_{i,j}-c) &   &{if \quad c \leqslant y_{i,j}\leqslant c+1}\\
  0 &   &{if \quad c+1 > y_{i,j}},
  \end{array} \right.
 \end{equation}
where $c$ is the channel index of the latent representation, $y_{i,j}$ denotes the $y$ value at the spatial position $(i,j)$, and $m_{i,j,c}$ represents the output of the importance map network at position $(i,j,c)$. 
The reason to introduce the importance map module is to reduce the bit rates for smooth regions. If all values of a channel are zeros, then no information will be sent for this channel during the encoding/decoding process.

In this paper, we develop an improved importance, as shown in Fig. \ref{BitAllocationSubnetworks}. Unlike \cite{Conditional}, which treats the first channel of the latent representation as the importance map, we design a separate network to extract importance map, which is more powerful. It also has the same number of channels as the latent representation, so that we can use element-wise product operator to merge them directly.

Our scheme consists of one convolution layer and three residual block modules. The output is denoted as $w$, which is then mapped to the range of $[-1,1]$ via a $\tanh()$ function and a $softsign()$ function, as shown in Eq. \ref{tanh_function} and \ref{sign_function}, where the output of the $\tanh()$ and $softsign()$ functions at position $(i,j,c)$ are denoted as $n(i,j,c)$ and $m(i,j,c)$ respectively.
  \begin{equation}\label{tanh_function}
   n_{i,j,c} = \tanh (w_{i,j,c})=\frac{e^{(w_{i,j,c})} - e^{-(w_{i,j,c})}}{e^{(w_{i,j,c})}+ e^{-(w_{i,j,c})}},
 \end{equation}
\begin{equation}\label{sign_function}
   m_{i,j,c} = softsign (n_{i,j,c})= \frac{n_{i,j,c}}{|n_{i,j,c}|+1}.
\end{equation}

\begin{algorithm}[!tp]
\caption{Channel Rate Allocation Algorithm} 
\hspace*{0.02in} {\bf Input:} 
The latent representation $\hat{y}$ ($1 \times W/16 \times H/16 \times 128$)\\
\hspace*{0.02in} {\bf Output:} 
output result
\begin{algorithmic}[1]
\State The state of each channel need to spend a bit to store
\State flag = [128//8] = 16 byte 
\For {i from 0 to 128}
\State Check if all elements in this channel are $0$
\If {$\hat{y}[0, :, :, i] == 0$}
\State flag[i] = 1
\Else
\State flag[i] = 0
\EndIf
\EndFor
\State \Return flag
\end{algorithmic}
\label{algorithm1}
\end{algorithm}

We then use element-wise product operator to merge $m$ and $y$ into $\tilde{y}$, i.e. $\tilde{y} \leftarrow y \odot m$. Next, $\tilde{y}$ is quantized and dequantized into the masked $\hat{y}$, which is sent to the entropy encoding. On the other hand, it is sent to the hyper networks to learn the probability distribution parameters of the  latent representation. 

As in \cite{Conditional}, we do not need to spend additional bits to store the importance map, as it can be automatically optimized by the loss function.

To better describe the channel rate allocation algorithm, we explain this processing in Algorithm \ref{algorithm1}. We assume that the number of channels $N$ is 128. We first use $\lceil 128//8 \rceil = 16$ bytes to represent the status of all channels. If all values of the channel are zeros, the flag of this channel is set to $1$, otherwise it is set to $0$. In the encoding and decoding processing, we only encode and decode the channels with non-zero coefficients.  The flags will be sent to the decoder. 

The importance map module can produce more zeros, save more bits, and reduce the encoding and decoding time. 

\subsection{2D Post-Quantization Filter (PQF)}

Quantization is an effective way to reduce the bit rate in image/video compression, but it introduces some quantization errors. In learned coding schemes, there is a mismatch between training and inference for quantization, because the training needs a differentiable quantization operator. 

To reduce the quantization error, we introduce a 2D post-quantization filter (PQF) stage, where a 3 × 3 convolution filter is applied to each feature map after entropy decoding. The coefficients  of the 3x3 filters are optimized to reduce the difference between the the dequantized $\hat{y}$ and the result before quantization, i.e., $\tilde{y}$. 

The following post-quantization loss function is added to the total cost function:
\begin{equation}
\begin{split}
L_{PQ}  = {\parallel F(\hat{y}; \omega,\theta) - \tilde{y} \parallel}^2,
\label{LPQ}
\end{split}
\end{equation}
where $F(\hat{y}; \omega, \theta)$ represents the 2D post-quantization filter with input $\hat{y}$ and parameters $\omega$ and $\theta$.

Our 2D PQF is motivated by the Sample Adaptive Offset (SAO) filter in H.265/HEVC and H.266/VVC \cite{SAO_chuamin}. However, our PQF is applied in the low-dimensional latent representation domain, and before decoder network. In the traditional video coding, the SAO filter is applied to the high-dimensional image domain after decoding. Besides, side information is needed in the SAO filter. In our case, the PQF is pre-trained  and does not need any side information.

\subsection{Loss Function}

The loss function of the entire system needs to consider both the number of output bits and the distortion of the reconstructed image.  To obtain different bit rates, we train several independent models with different values of the Lagrange multiplier $\lambda$. The PQF cost function in Eq. \ref{LPQ} is also included in the total cost function. The entire loss function is as follows.
\begin{equation}\label{total_loss}
\begin{aligned}
 L =  \lambda D(x,\hat{x})&+H(\hat{y})+H(\hat{z}) + \lambda_{1} L_{PQ}, \\
      H(\hat{y}) &=  E [-\log_{2}(P_{\hat{y}|\hat{z}}(\hat{y}|\hat{z}))],\\
      H(\hat{z}) &=  E [-\log_{2}(P_{\hat{z}}(\hat{z}))],
\end{aligned}
\end{equation}
where $D(x, \hat{x})$ is the reconstruction error between origin image $x$ and the reconstructed image $\hat{x}$, which can be measured by mean squared error (MSE) or MS-SSIM. $H(\hat{y})$, $H(\hat{z})$ are the entropies of the core latent representation and hyper representations, which represent their bit rates. $\lambda_{1}$ is another Lagrangian parameter for the 2D PQF.

\section{Implementation Details}
\label{Implementation}

\subsection{Datasets}

We select the CLIC dataset \cite{CLIC} and LIU4K dataset \cite{LIU_dataset} to train the proposed models. All training images of the two datasets are rescaled to a resolution of $2000 \times 2000$. We utilize data augmentation technologies (i.e., randomly rotation and scaling) to collect 81,650 training images with a resolution of $384 \times 384$. These training images are saved as lossless PNG images.

The test datasets are the Kodak PhotoCD dataset \cite{Kodak} and Tecnick dataset \cite{Tecnick}. The Kodak dataset consists of 24 images with resolution of $768 \times 512$ or $512 \times 768$. The Tecnick dataset has 40 high-quality images with a resolution of $1200 \times 1200$.

\begin{table*}[!t]
\centering %
\caption{Comparisons of encoding and decoding time, BD-Rate, and model sizes.}
\begin{tabular}{|c|c|c|c|c|c|c|}
\hline
\textbf{Dataset}& \textbf{Method} & \textbf{Encoding time} & \textbf{Decoding Time} & \textbf{BD-Rate} &\textbf{Model size(Low bit rates) }&\textbf{Model size(High bit rates)} \\ 
\hline
\multirow{7}{*}{Kodak}  & VVC  & 402.27s& 0.607s& 0.0 & 7.2 MB & 7.2MB\\ \cline{2-7} 
                         & Lee2019 \cite{Lee_2020}  &10.721s& 37.88s& 17.0\% & 123.8 MB & 292.6MB\\ \cline{2-7} 
                         & Hu2021 \cite{Hu_2021}  &35.7187s& 77.3326s& 11.1 \% & 84.6 MB & 290.9MB\\ \cline{2-7} 
                         & Cheng2020 \cite{cheng2020}  &20.89s& 22.14s& 4.8 \% & 50.8 MB & 175.18MB\\
                         \cline{2-7} 
                         & Chen2021 \cite{chen2021}  &402.26s& 2405.14s& 8.6 \% & 200.99 MB & 200.99MB\\
                         \cline{2-7} 
                         & GLLMM \cite{GLLMM}  &467.90s& 467.90s& -3.13\% & 77.08 MB & 241.03MB\\ \cline{2-7} 
                         & \textbf{Ours}  &\textbf{22.62s}& \textbf{23.512s}& \textbf{-2.12\%} & \textbf{54.06MB} &  \textbf{181.3MB}\\ 
                         \hline
                         
\multirow{7}{*}{Tecnick}  & VVC  & 700.59s& 1.49s& 0.0 & 7.2 MB & 7.2MB\\ \cline{2-7} 
                         & Lee2019 \cite{Lee_2020} &54.8s& 138.81s& 30.3 \%  & 123.8 MB & 292.6MB\\ \cline{2-7} 
                         & Hu2021 \cite{Hu_2021} &84.035s& 271.50s& 16.8 \% & 84.6 MB & 290.9MB\\ \cline{2-7} 
                         & Cheng2020 \cite{cheng2020} &59.48s& 71.71s& 6.5 \% & 50.8MB &175.18MB\\
                         \cline{2-7} 
                         &Chen2021 \cite{chen2021}  &62.33s& 4503.25s& 8.6 \% & 200.99 MB & 200.99MB\\
                         \cline{2-7} 
                         & GLLMM \cite{GLLMM}   &1233.05s& 1245.05s& -4.25\%& 77.08 MB & 241.03MB\\ \cline{2-7} 
                         &\textbf{ Ours}  &\textbf{45.90}s& \textbf{47.705s}& \textbf{-3.44\%} & \textbf{54.06 MB} & \textbf{181.3MB}\\ 
                         \hline
\end{tabular}
\label{runing_time}
\end{table*}

\subsection{Parameter Settings}

Both mean squared error (MSE) and MS-SSIM are considered as distortion metrics to optimize our models. When optimized for MSE, the parameter $\lambda$ is chosen from the set $\{0.0016,0.0032,0.0075,0.015, 0.023, 0.03, 0.045\}$. Each trains an independent model for a bit rate. The number of channels $N$ in the latent representation is set to 128 for the first three cases, and is increased to 256 for the last four cases. When the MS-SSIM metric is used, the parameter $\lambda$ is set to 12, 40, 80 and 120. The value of $N$ is set to 128 for the first two cases, and 256 for the other two cases. Each model is trained for $1.5 \times 10^{6}$ iterations. The Adam solver with a batch size of 8 is adopted. The learning rate is set to $1 \times 10^{-4}$ in the first 750,000 iterations, and reduced by 0.5 after every 100,000 iterations in the last 750,000 iterations. 
We set $ \lambda_{1} $ to 1 in the first 20,000 iterations so that the PQF can regulate the decoded $ \hat{y} $ to make it more close to the result before quantization. Then we set $ \lambda_{1} $ to 0 in the remaining iterations to remove the regulation term and let the whole framework train only with bitrate terms and distortion term in order to obtain optimal compression performance.

\section{Experimental Results}
\label{Experiment}

In this section, we compare the proposed method with the state-of-the-art learned image compression methods and classical methods in terms of encoding/decoding complexity, PSNR, and MS-SSIM metrics. The MS-SSIM is measured in decibels by $-10 \log_{10}(1 - MS\textnormal{-}SSIM)$. The learned methods include GLLMM \cite{GLLMM}, Chen2021 \cite{chen2021}, Ma2021 \cite{Ma_2022_PAMI}, Cheng2020 \cite{cheng2020}, Hu2021 \cite{Hu_2021}, and Lee2019 \cite{Lee_2020}. The classical methods include the latest H.266/VVC-Intra (4:4:4) (VTM 8.0) \cite{vvc}, VVC-Intra (4:2:0), BPG (H.265/HEVC-Intra), WebP, JPEG 2000, and JPEG.

\begin{figure*}[!thp]
\centering
\subfigure
{\begin{minipage}[t]{0.5\linewidth}
\centering
\includegraphics[width=\columnwidth]{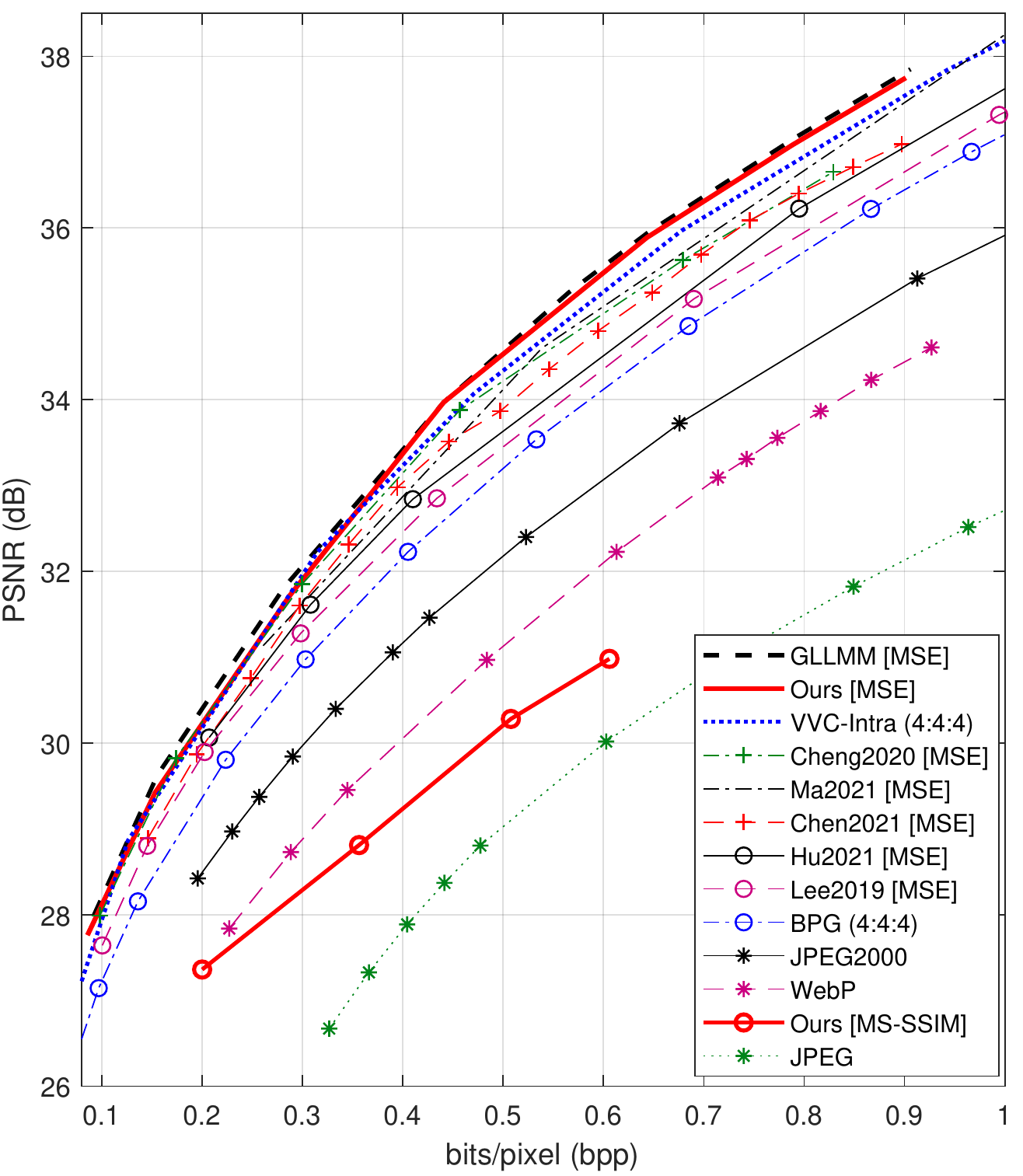}
\end{minipage}
}%
\subfigure
{\begin{minipage}[t]{0.5\linewidth}
\centering
\includegraphics[width=\columnwidth]{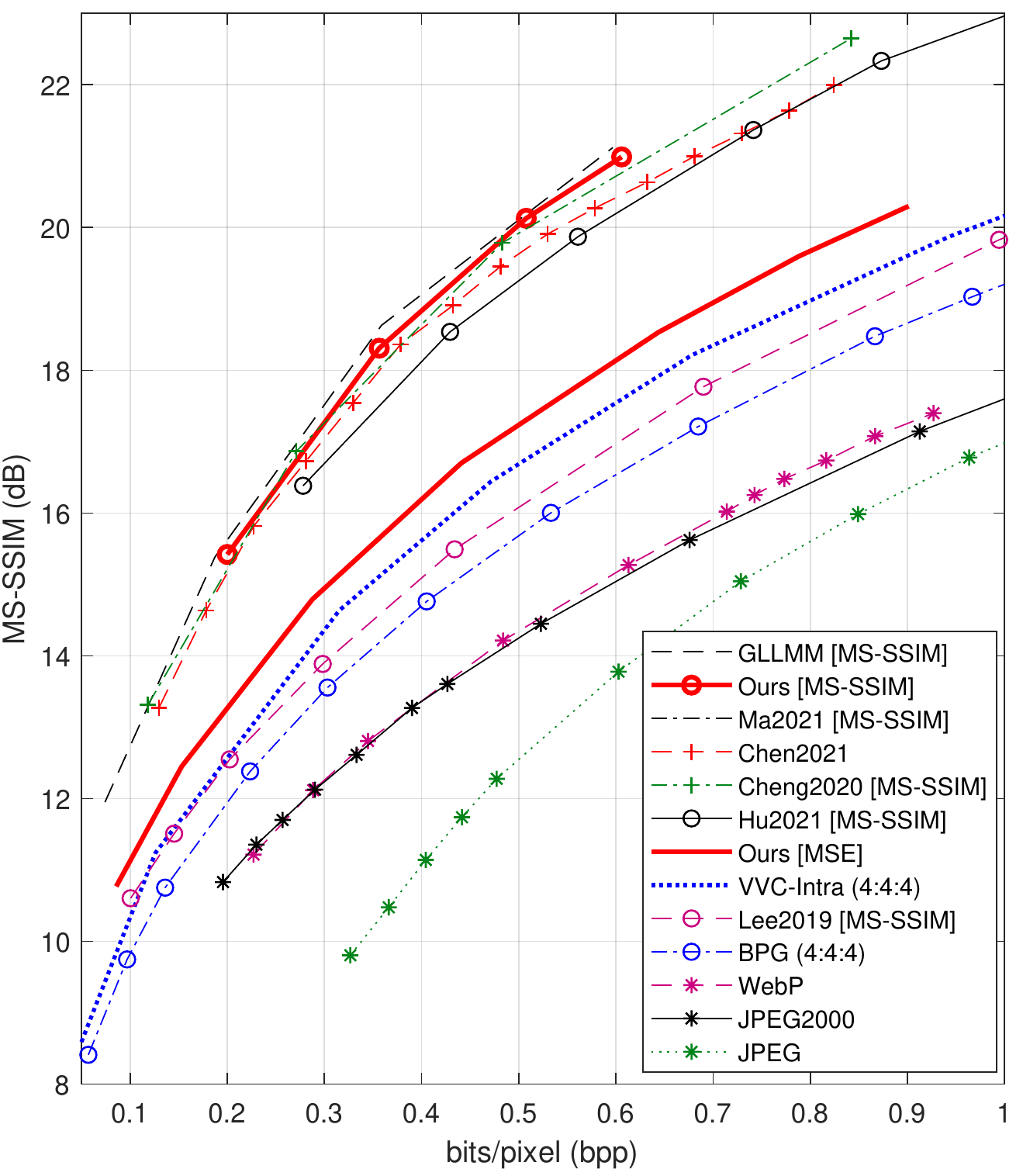}
\label{kodak_SSIM}
\end{minipage}
}%
\centering
\caption{The R-D curves of different methods on the Kodak dataset when optimized for PSNR and MS-SSIM respectively.}
\label{fig:kodak}
\end{figure*}

\begin{figure*}[!thp]
\centering
\subfigure
{\begin{minipage}[t]{0.5\linewidth}
\centering
\includegraphics[width=\columnwidth]{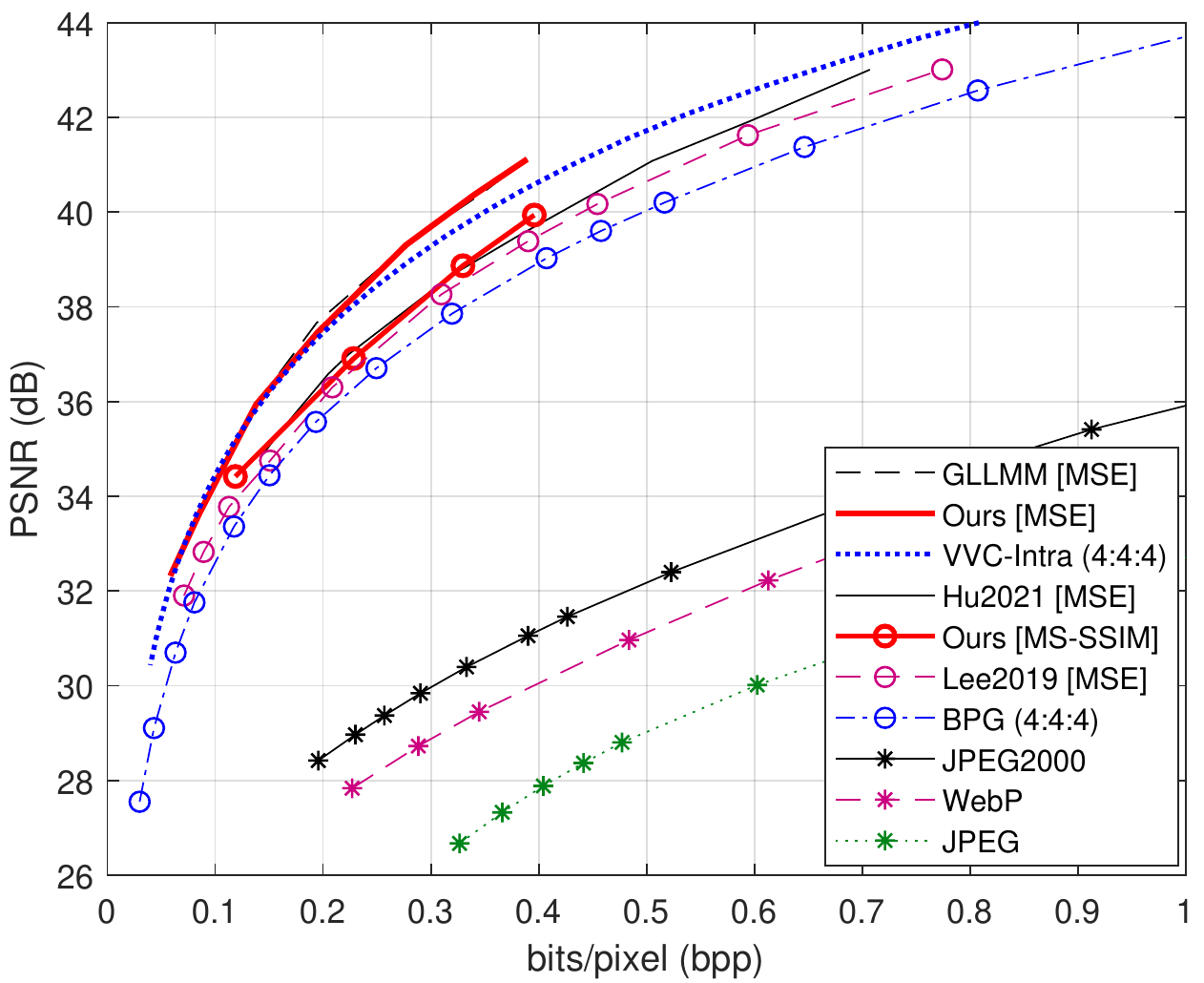}
\end{minipage}
}%
\subfigure
{\begin{minipage}[t]{0.5\linewidth}
\centering
\includegraphics[width=\columnwidth]{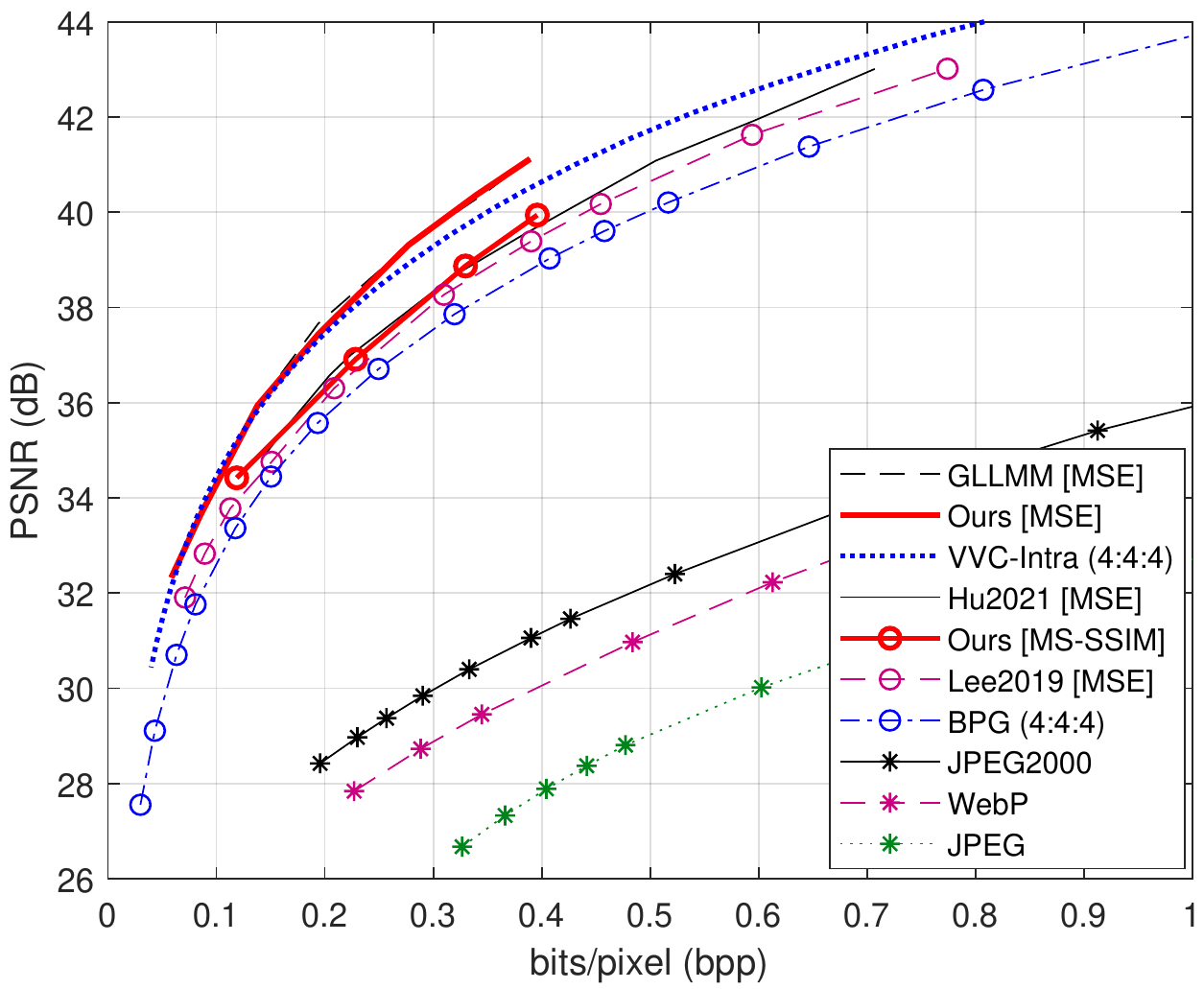}
\end{minipage}
}%
\centering
\caption{The R-D curves of different methods on the Tecnick dataset when optimized for PSNR and MS-SSIM respectively.}
\label{fig:tecnick}
\end{figure*}

\subsection{Encoding and Decoding Complexity}

Table \ref{runing_time} compares the complexities of different methods. Since VVC, Hu2021 \cite{Hu_2021}, Cheng2020 \cite{cheng2020} only run on CPU, we evaluate the running time of different methods on an 2.9GHz Intel Xeon Gold 6226R CPU. The average time over all images in each dataset is used. The average model sizes at low bit rates and high bit rates are also reported. In addition, we use VVC as the anchor scheme, and calculate the BD-Rate saving of other methods with respect to VVC \cite{BDRate}.

It can be seen from Table \ref{runing_time} that compared to the state-of-the-art GLLMM method \cite{GLLMM}, the proposed scheme is about 17 times faster in both encoding and decoding, and the BD-Rate only loses about $1\%$ and $0.8\%$ for the two datasets. Our model size is also smaller than \cite{GLLMM}, and is comparable to Cheng2020 \cite{cheng2020}, but our BR-Rate is $6.92\%$ and $9.94\%$ better than \cite{cheng2020}. Compared to VVC, our encoding is about 14 times faster, but the decoding is much slower. Our BD-Rate is $2.12\%$ and $3.44\%$ better than VVC. Therefore the proposed scheme achieves the new state of the art in learned image coding when considering both the complexity and performance.

\subsection{Rate-Distortion Performances}

The average R-D curves of the 24 Kodak images are shown in Fig. \ref{fig:kodak}. When optimized for PSNR, GLLMM \cite{GLLMM} achieves the best result, even higher than VVC (4:4:4), especially at high rates. When the bit rate is lower than 0.4 bpp, our method is about 0.1 dB lower than GLLMM, but above 0.4 bpp, our method obtains the same performance as GLLMM. Our method is similar to Cheng2020 \cite{cheng2020} when the bit rate is lower than 0.3 bpp. After that, we can get better performance.

When optimized for MS-SSIM, as in Fig.\ref{kodak_SSIM}, GLLMM, Cheng2020, and our method have similar performance, and is about 3 dB higher than our PSNR-optimized method, but even the latter is better than VVC.

Fig. \ref{fig:tecnick} shows the R-D performances on the Tecnick dataset. Some methods are not included because we do not have their results. It can be observed that the proposed method achieves almost identical result to GLLMM in both PSNR and MS-SSIM, and is also better than VVC in the more challenging case of PSNR. Our method is also more than 1dB better than other learned methods at high rates.

\begin{figure}[!t]
\centering
\subfigure
{\begin{minipage}[t]{1\linewidth}
\centering
\includegraphics[width=\columnwidth]{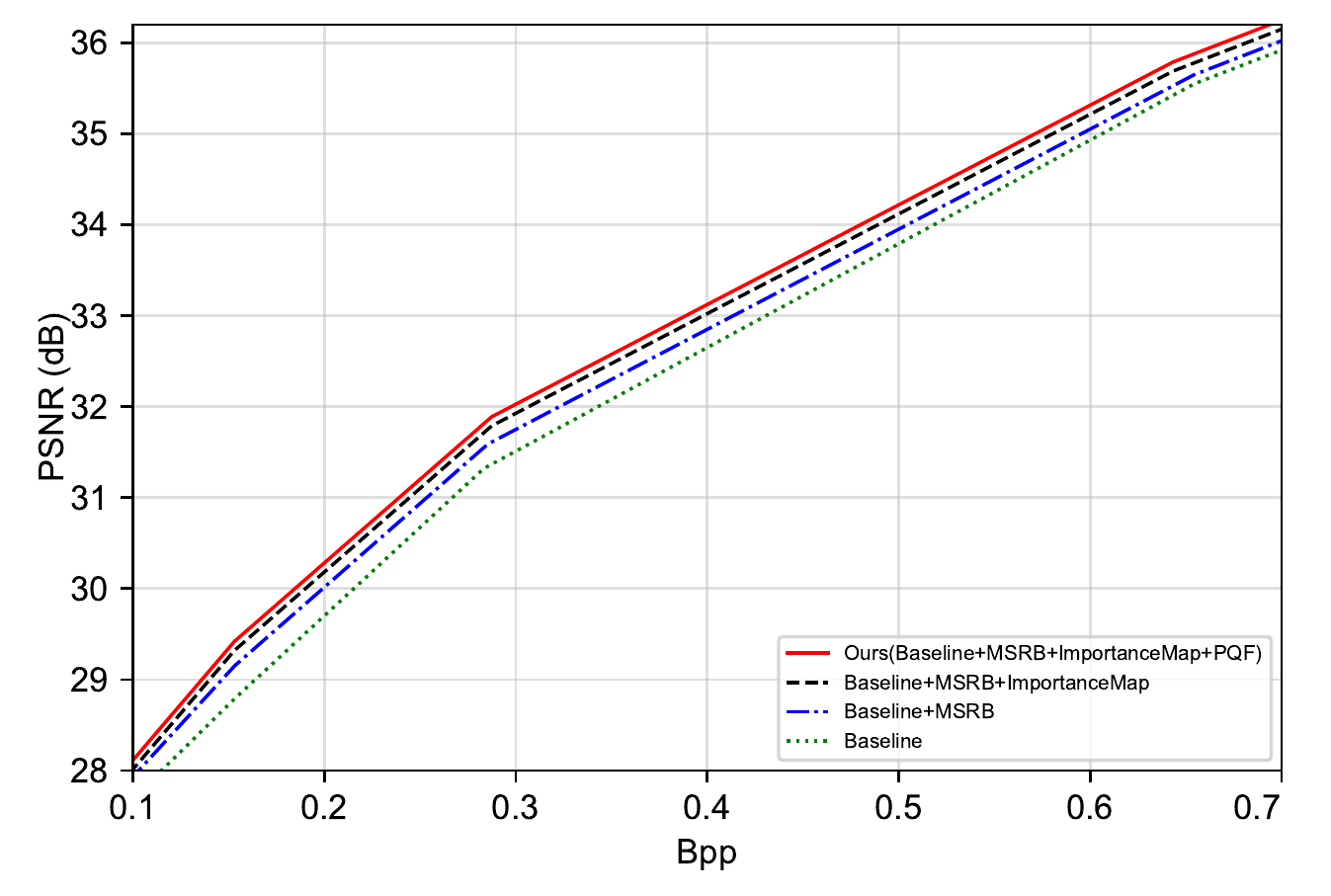}
\end{minipage}
}%
\caption{The performance of various components in the proposed scheme.}
\label{fig:different_modules}
\end{figure}

\subsection{Ablation Studies}
\label{Ablation}
\subsubsection{The Effectiveness of Various Components }

We first show in Fig. \ref{fig:different_modules} an ablation study to demonstrate the effectiveness of various components in our proposed scheme in Fig. \ref{networkstructure}. The parameters are the same as our method when optimized for PSNR. We first remove the MSRB, importance map, and the PQF from the proposed method. The remaining network is used as the baseline. We then add the three components sequentially. 

Compared to the baseline, \textit{Baseline+MSRB} improves the R-D  performance about 0.15-0.2 dB at the same bit rate. Compared to \textit{Baseline+MSRB}, the \textit{Baseline+MSRB+ImportanceMap} improves the R-D  performance about 0.1-0.15 dB at the same bit rate. Compared to \textit{Baseline+MSRB+ImportanceMap}, the proposed full method \textit{Ours (Baseline+MSRB+ImportanceMap+QPF)} in this paper has a further improvement of 0.1 dB. It can be observed that MSRB contributes the most, followed by the importance map and the PQF, especially at low rates. 

\begin{table}[!tp]
\caption{Comparison of our importance map to that in \cite{Conditional}.}
\begin{center}
  \begin{tabular}{ccccc}
  \hline
  \textbf{Dataset}& \textbf{Scheme}  & \textbf{bit rate}& \textbf{ MS-SSIM}& \textbf{PSNR}\\
  \hline
  \textbf{Kodak}     & \cite{Conditional}  & 0.1551  & 0.9423  & 29.28\\
  \textbf{Kodak}     & Ours                           & 0.1531  & 0.9432  & 29.42\\
  \textbf{Tecnick}  & \cite{Conditional}   & 0.0901  & 0.9612  & 33.52\\
  \textbf{Tecnick}  & Ours                            & 0.0855  & 0.9628  & 33.65\\
  \hline
  \textbf{Kodak}    & \cite{Conditional}   & 0.9052  & 0.9900  & 37.62\\
  \textbf{Kodak}    & Ours                            & 0.9013  & 0.9907  & 37.75\\
  \textbf{Tecnick}  & \cite{Conditional}   & 0.3989  & 0.9923  & 23.85\\
  \textbf{Tecnick}  & Ours                            & 0.3959  & 0.9961  & 24.01\\
  \hline
\end{tabular}
\label{important_map_compare}
\end{center}
\end{table}

\subsubsection{Comparison with Previous Importance Map}
Table \ref{important_map_compare} compares our importance map network with that in \cite{Conditional}, by replacing our importance map with the latter in our structure. We train two models at a low bit rate and a high bit rate, respectively, optimized for PSNR. At the low bit rate, $N=128$ and $\lambda=0.0032$. At the high bit rate, $N=256$ and $\lambda=0.06$.

As shown by Table \ref{important_map_compare}, our importance map network achieves about 0.15 dB higher PSNR than \cite{Conditional}, although sometimes the bit rate of our scheme is even lower.

\begin{figure}[!th]
\subfigure[]{
\begin{minipage}[t]{0.45\linewidth}
\includegraphics[scale=0.16]{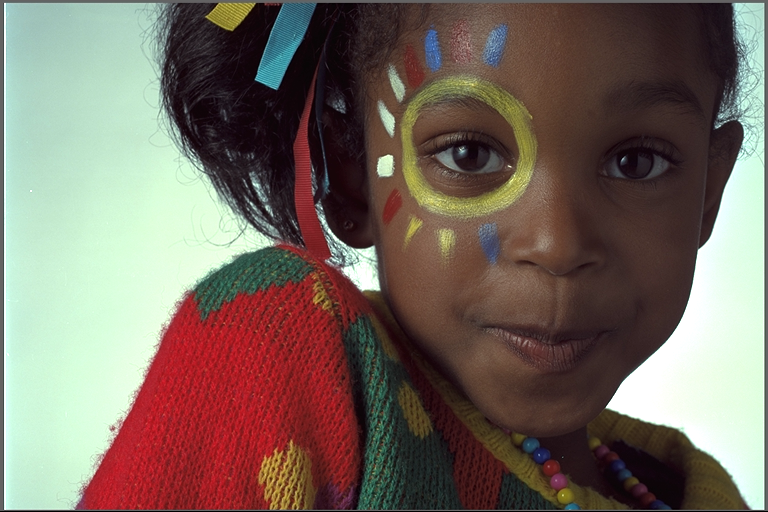}
\end{minipage}
}%
\subfigure[]{
\begin{minipage}[t]{0.45\linewidth}
\includegraphics[scale=1.3]{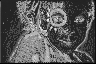}
\end{minipage}
}%
\caption{(a) An origin image in the Kodak dataset. (b) The importance map obtained by our method at 0.085 bpp.}
\label{fig:importance_map}
\end{figure}

\begin{figure}[!thp]
\centering
{\begin{minipage}[t]{1\linewidth}
\centering
\includegraphics[width=\columnwidth]{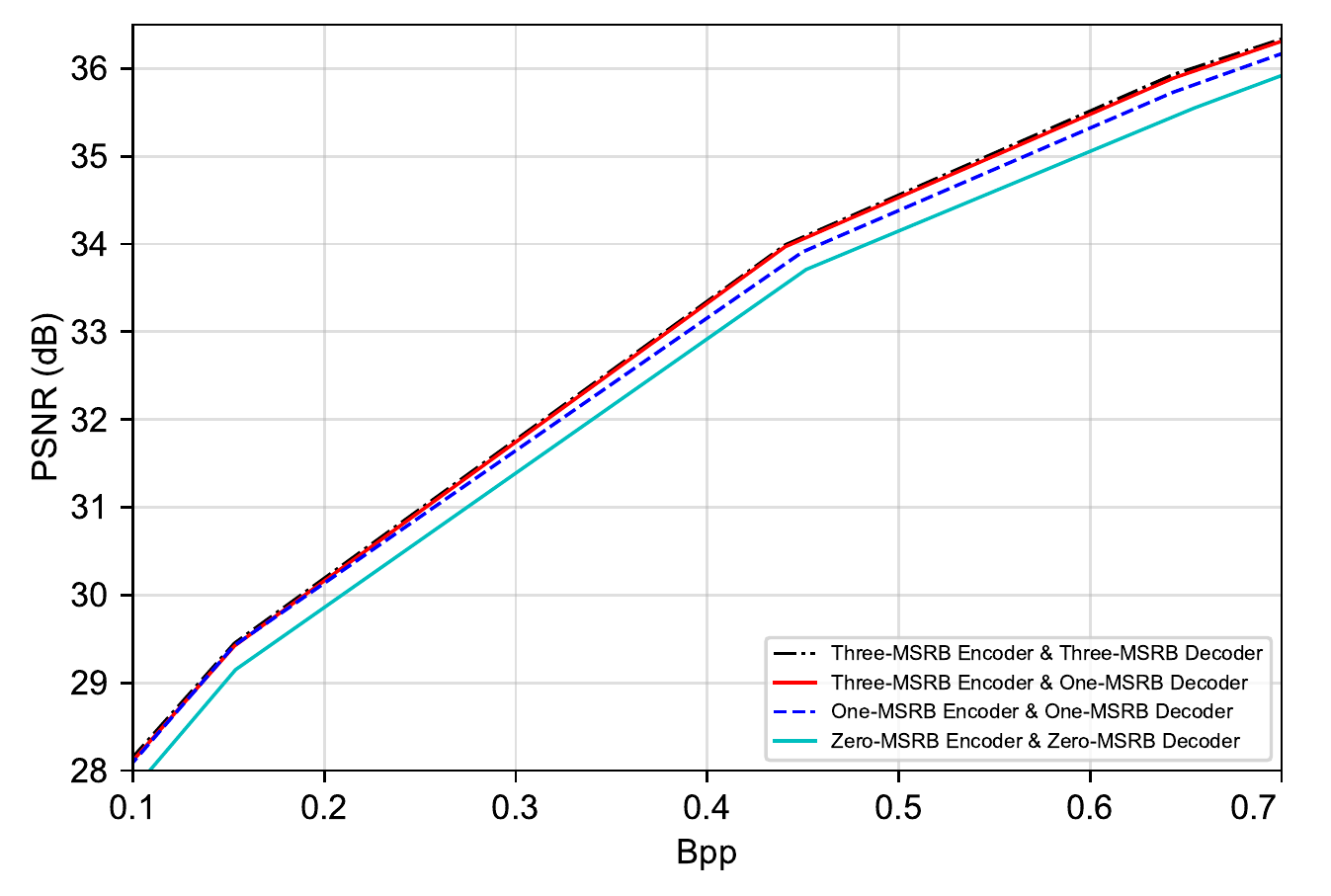}
\end{minipage}
}%
\centering
\caption{The performance of different MSRBs in the core encoder and decoder.}
\label{fig:MSRB_encoder}
\end{figure}

Fig. \ref{fig:importance_map} shows an example of the importance map obtained by our method for an image in the Kodak dataset. Latent representations with higher values of importance (brighter) will be allocated more bits than those with lower values (darker). It can be seen that the importance map allocates more bits to complex and salient regions, and less bits to smooth regions.


\begin{table*}[!thp]
\caption{The comparison of different decoding methods.}
\begin{center}
  \begin{tabular}{ccccccc}
  \hline
 \textbf{Name} & \textbf{Bit rates} & \textbf{Zeros Channels} & \textbf{Total Channels(N)}  & \textbf{Decoding time(Ours)} & \textbf{Decoding time(Full)} &\textbf{ Reduce Rate (Decoding)}\\
  \hline
  \textbf{Kodak} & Low bit rates &79 & 128  & 23.5196 s & 47.7051s & 50.69\%\\
  \textbf{Kodak} & High bit rates &134 & 256  &178.012 s  &396.32 s   & 55.08\%\\
  \hline
  \textbf{Tecnick} & Low bit rates &62 & 128   & 47.7051s  & 96.13 s & 50.37 \%\\
  \textbf{Tecnick}   & High bit rates &89 & 256  & 333.99 s  & 570.48 s &41.45\%\\
  \hline
\end{tabular}
\label{Channel_algorithem}
\end{center}
\end{table*}

\begin{table}[!thp]
\caption{The comparison of MSRB and the CRB in \cite{GLLMM}. }
\begin{center}
  \begin{tabular}{cccc}
  \hline
 \textbf{Module} & \textbf{Bit rate}  & \textbf{PSNR (dB)} & \textbf{MS-SSIM}\\
  \hline
  \textbf{CRB \cite{GLLMM}}               & 0.1602 & 29.46 & 0.9424 \\
  \textbf{OriginalMSRB \cite{MSRB_ECCV18}} & 0.1547 & 29.22 & 0.9415 \\
  \textbf{MSRB (Ours)}                    & \textbf{0.1531} &\textbf{29.48}  &\textbf{0.9427}\\
  \hline
  \textbf{CRB \cite{GLLMM}} & 0.8037  &36.80 &0.9886  \\
  \textbf{OriginalMSRB \cite{MSRB_ECCV18}} & 0.7896  &36.80 &0.9884  \\
 \textbf{MSRB (Ours)} & \textbf{0.7886} &\textbf{36.97} & \textbf{0.9890}  \\
  \hline
\end{tabular}
\label{MSRBvsCRB}
\end{center}
\end{table}

The importance map network can facilitate to produce more zeros in latent representation, which cooperates with the channel allocation algorithm in Algorithm \ref{algorithm1} to effectively shorten the decoding time. We test the channel allocation algorithm in Kodak and Tecnick  datasets on decoding time. We first count the number of channels whose values are all 0, denoted \textbf{Zeros Channels} in Table \ref{Channel_algorithem}. The \textbf{Decoding time(ours)} in Table \ref{Channel_algorithem} represents we only encode and decode the channels with non-zero coefficients. The \textbf{Decoding time(Full)} in Table \ref{Channel_algorithem} represents we encode and decode the whole channels. The \textbf{Reduce Rate (Decoding)} in Table \ref{Channel_algorithem} represents \textbf{Decoding time(ours)} can reduce decoding time rate compared with  \textbf{Decoding time(Full)}. It is concluded that the decoding time (ours) can be reduced by 40\% at low bit rates and high bit rates.

\subsubsection{Comparison of MSRB, Original MSRB and CRB}
Table \ref{MSRBvsCRB} compares our improved MSRB with the original MSRB in \cite{MSRB_ECCV18} and the concatenated residual block (CRB) in \cite{GLLMM} in terms of PSNR and MS-SSIM metrics. We respectively replace the improved MSRB by CRB and original MSRB in our scheme. Other structures remain the same. We train two models at a low bit rate and a high bit rate. As shown by the table \ref{MSRBvsCRB}, our proposed MSRB achieves better performance than  CRB and original MSRB at low and high bit rates on both PSNR and MS-SSIM metrics. 

\subsubsection{Number of MSRBs in Encoder and Decoder}
As shown in Fig. \ref{fig:MSRB_encoder}, we find that the performance hardly degrades when we use less number of MSRBs in the decoder, which allows us to reduce the complexity of the overall network. The reason is that the main role of the decoder is to obtain the reconstructed image. A well learned feature tensor is sufficient to represent the image and can be reconstructed by a less complex decoder.


A natural question is if we can reduce the number of MSRBs in the encoder. To show this, we use only one MSRB in the encoder, and compare with the proposed encoder with three MSRBs. The result on the Kodak dataset is shown in Fig. \ref{fig:MSRB_encoder}. At the low bit rate, the one-MSRB encoder can achieve similar performance to the three-MSRB encoder. However, when the bit rate increases, the performance of the one-MSRB encoder will drop about 0.1-0.15 dB compared to the three-MSRB encoder. The reason is that at the low bit rates, the network capacity with one MSRB is large enough to extract compact and efficient latent representation, but at high bit rates, one MSRB does not have high learning capability and more MSRBs are needed. But if we remove the MSRB from encoder and decoder networks, the image compression will drop 0.2-0.3 dB regarding PSNR at a range of bit rates.

\subsubsection{Impact of the Attention module}

\begin{figure}[!tp]
\centering
{\begin{minipage}[t]{1\linewidth}
\centering
\includegraphics[width=\columnwidth]{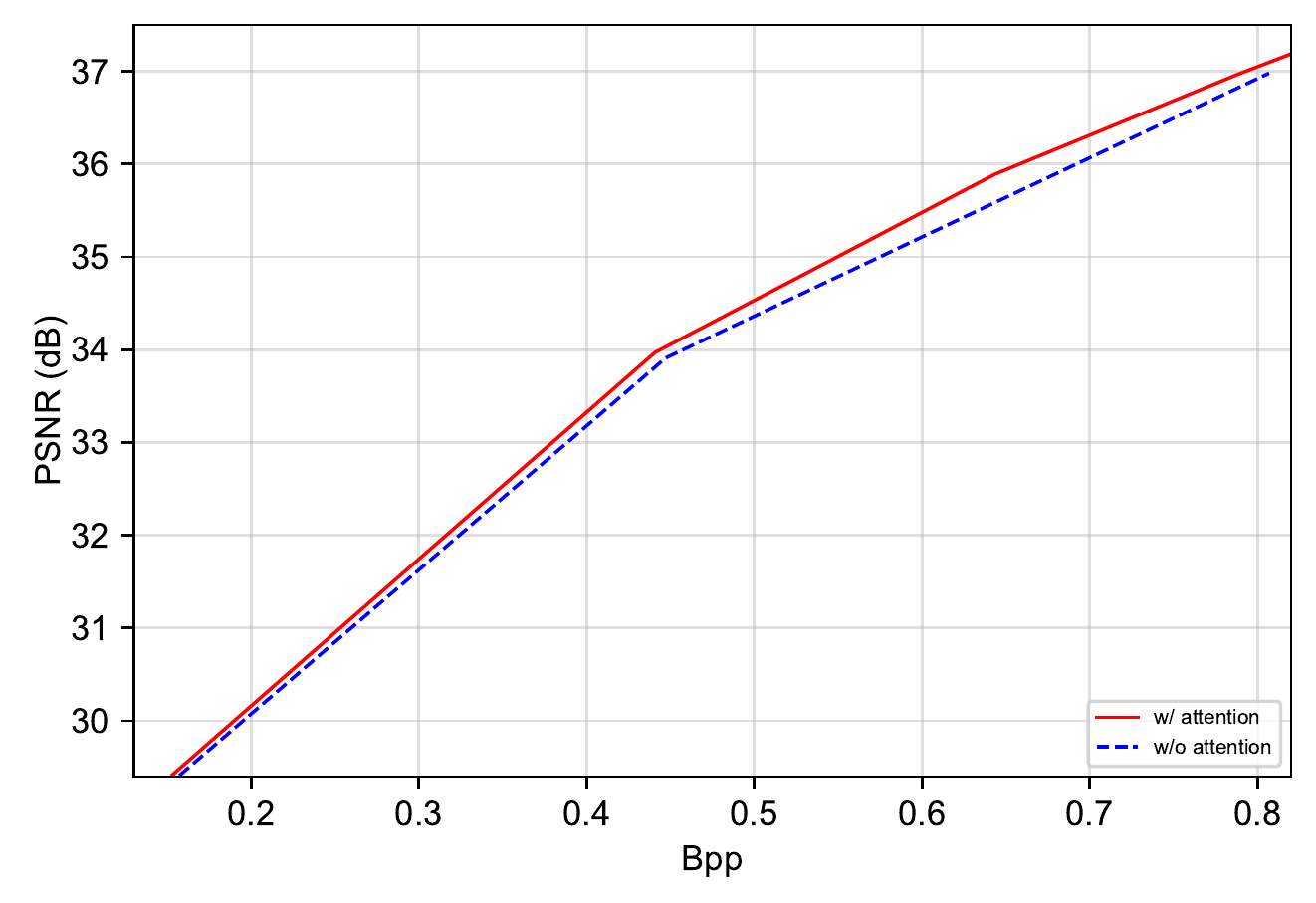}
\end{minipage}
}%
\centering
\caption{The performance of attention module in \cite{cheng2020}.}
\label{fig:attension_module}
\end{figure}

Fig. \ref{fig:attension_module} shows the effectiveness of the attention module. Different from the importance map network where the importance is leveraged for bit allocation, the attention module \cite{cheng2020} is used to extract informative regions in the encoder and decoder, thus improving the final reconstruction. The result \textit{w/o attention} represents that we remove the attention modules from our network architecture. The compression performance from \textit{w/o attention} drops 0.1-0.15 dB in term of PSNR compared to that from \textit{w/ attention}. Therefore, we add the attention module in our network structure to enhance the compression performance.

\subsection{Qualitative Results}

\begin{figure*}[!tph]
\centering
\subfigure[Original]{
\begin{minipage}[t]{0.33\linewidth}
\centering
\includegraphics[scale=0.4]{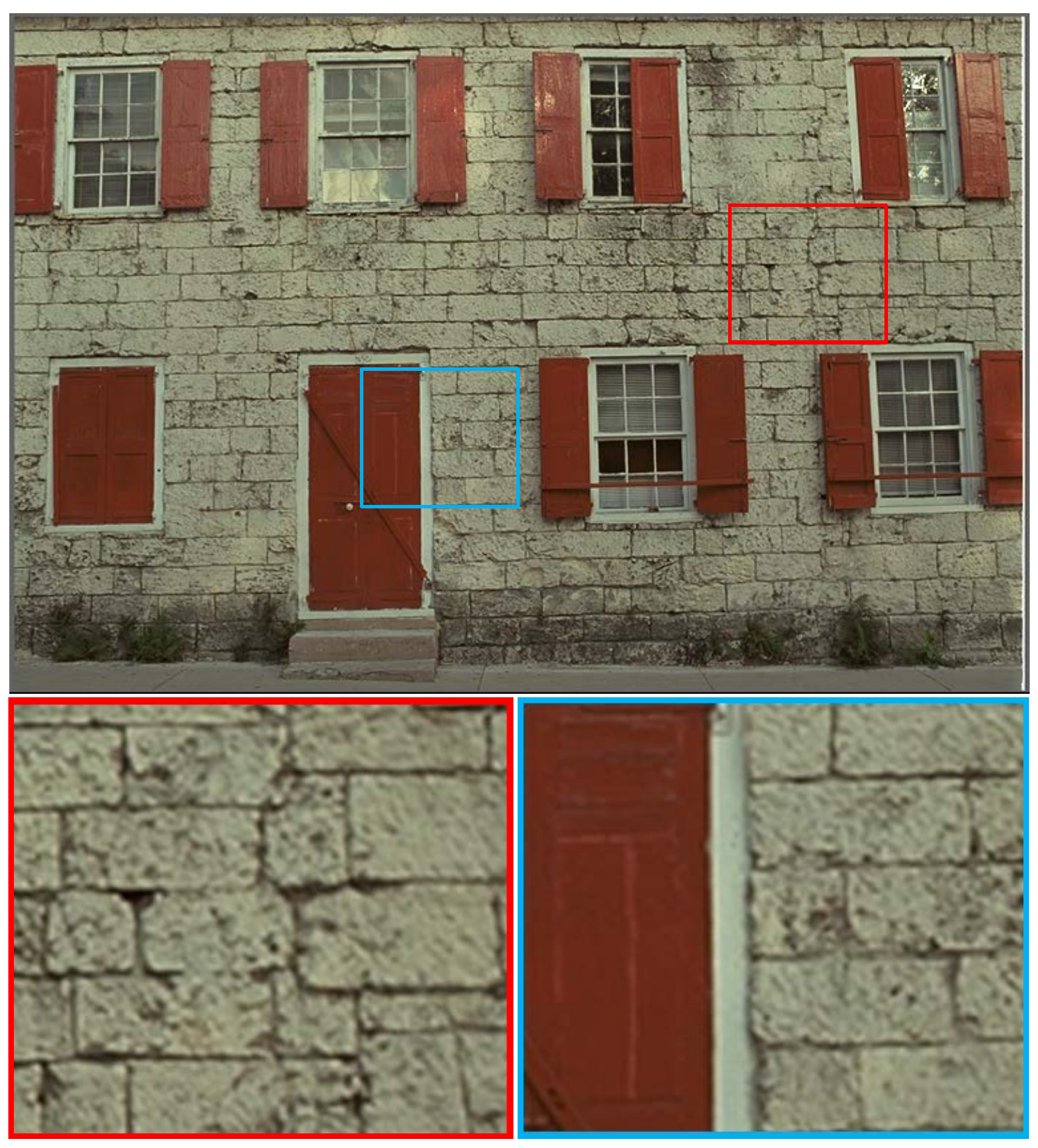}
\end{minipage}
}%
\subfigure[JPEG(0.239/21.01/0.777)]{
\begin{minipage}[t]{0.33\linewidth}
\centering
\includegraphics[scale=0.4]{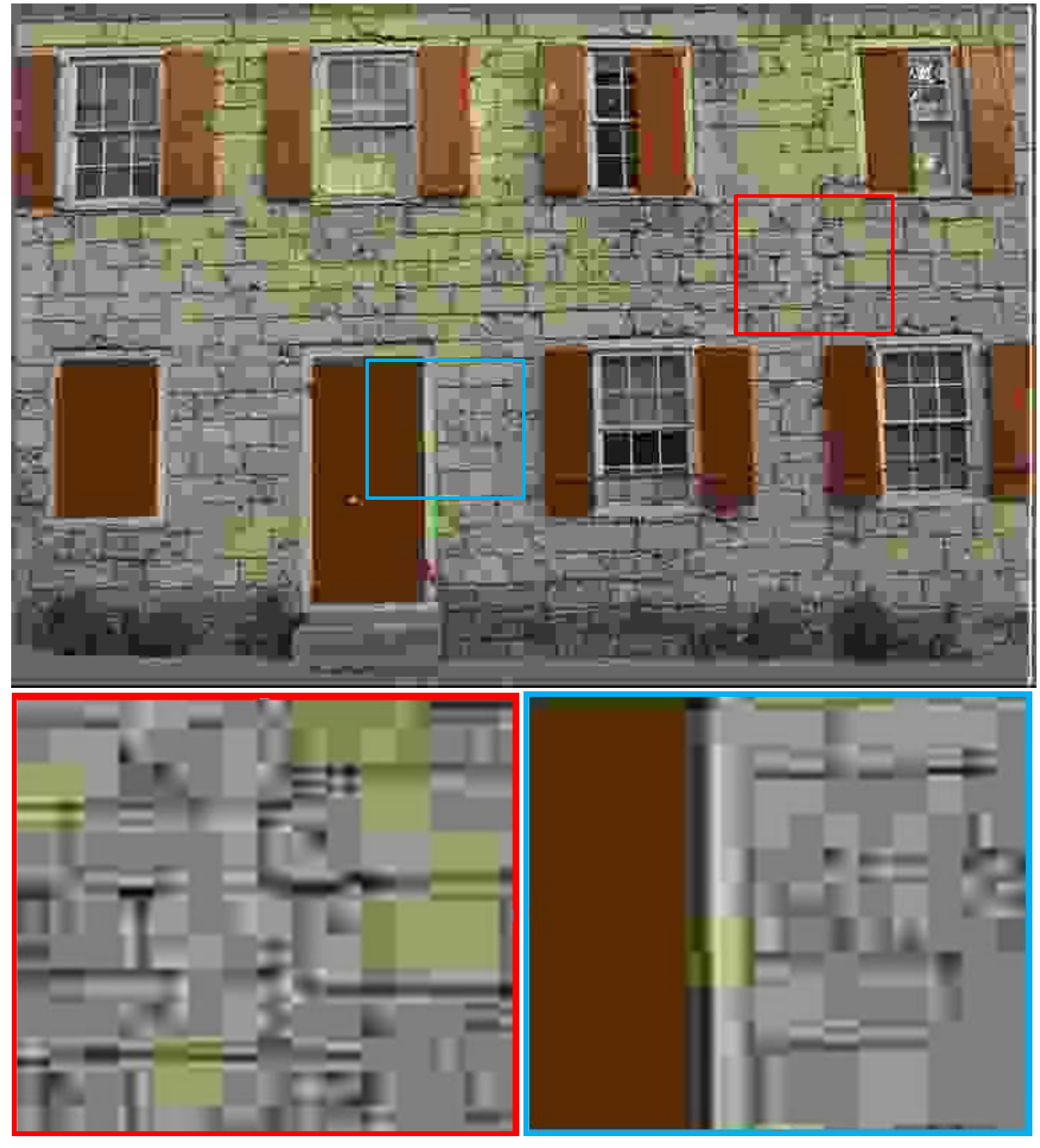}
\end{minipage}
}%
\subfigure[JPEG2000(0.234/23.84/0.895)]{
\begin{minipage}[t]{0.33\linewidth}
\centering
\includegraphics[scale=0.4]{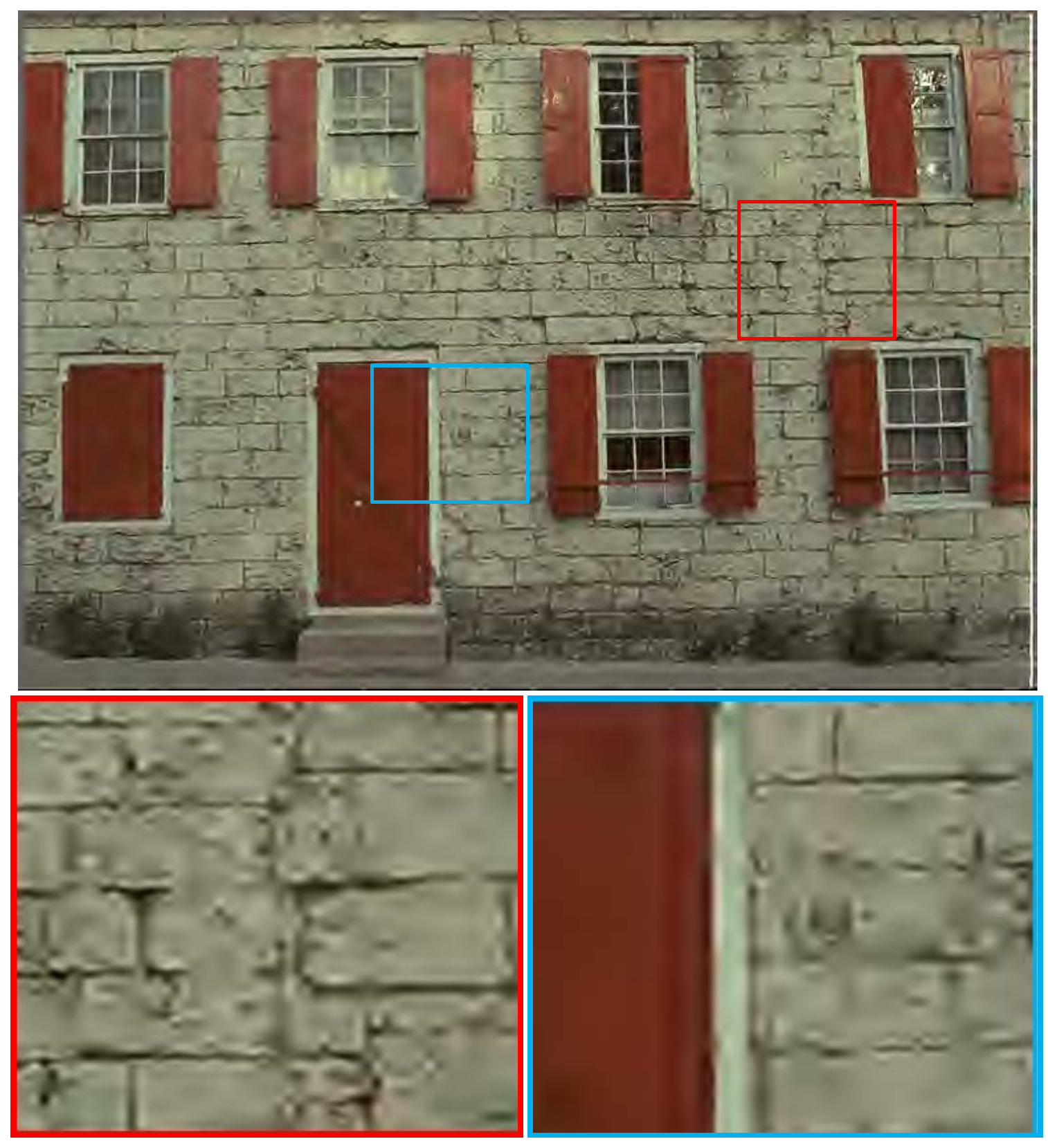}
\end{minipage}
}%

\subfigure[BPG(0.235/26.08/0.925]{
\begin{minipage}[t]{0.33\linewidth}
\centering
\includegraphics[scale=0.4]{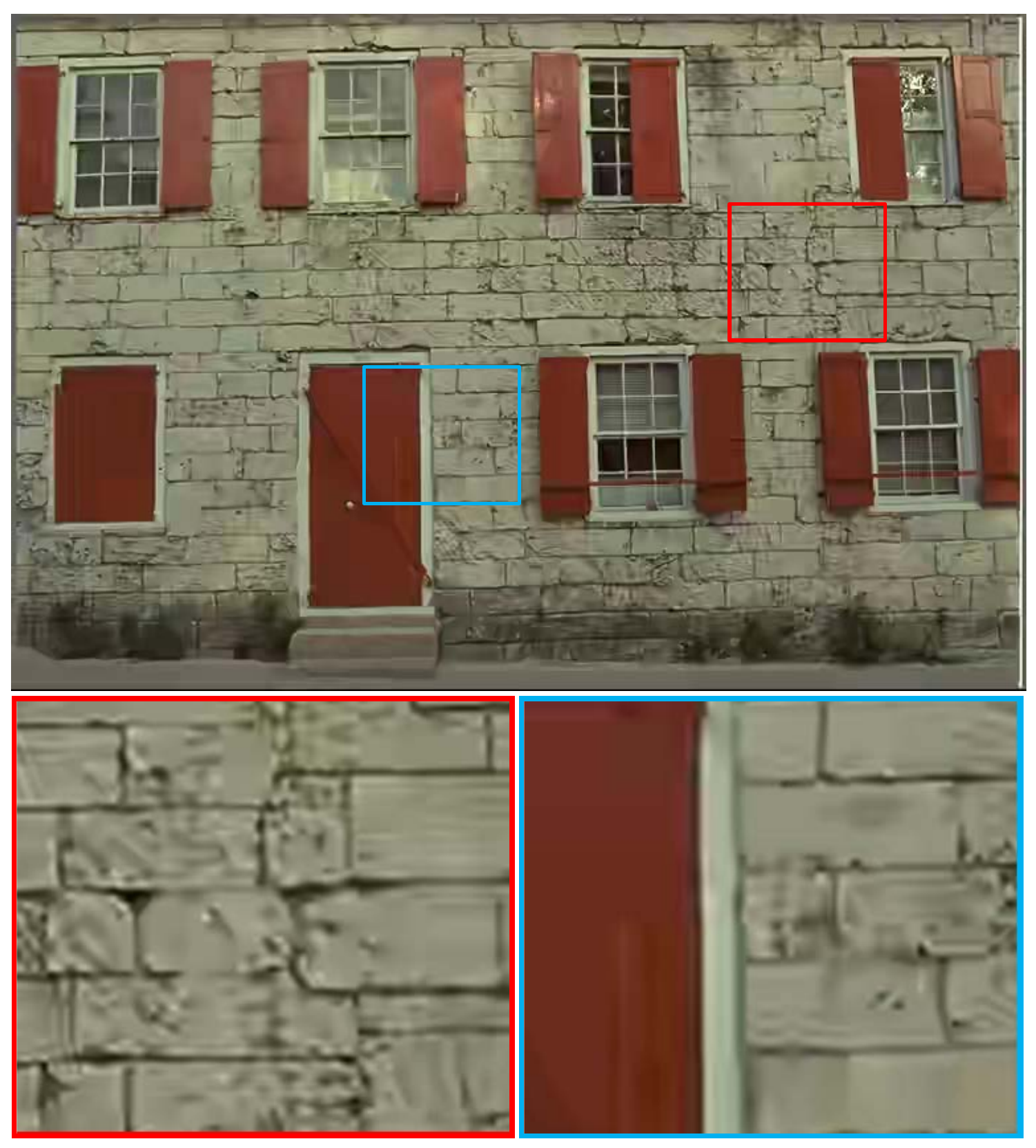}
\end{minipage}
}%
\subfigure[VVC(0.231/26.73/0.934)]{
\begin{minipage}[t]{0.33\linewidth}
\centering
\includegraphics[scale=0.4]{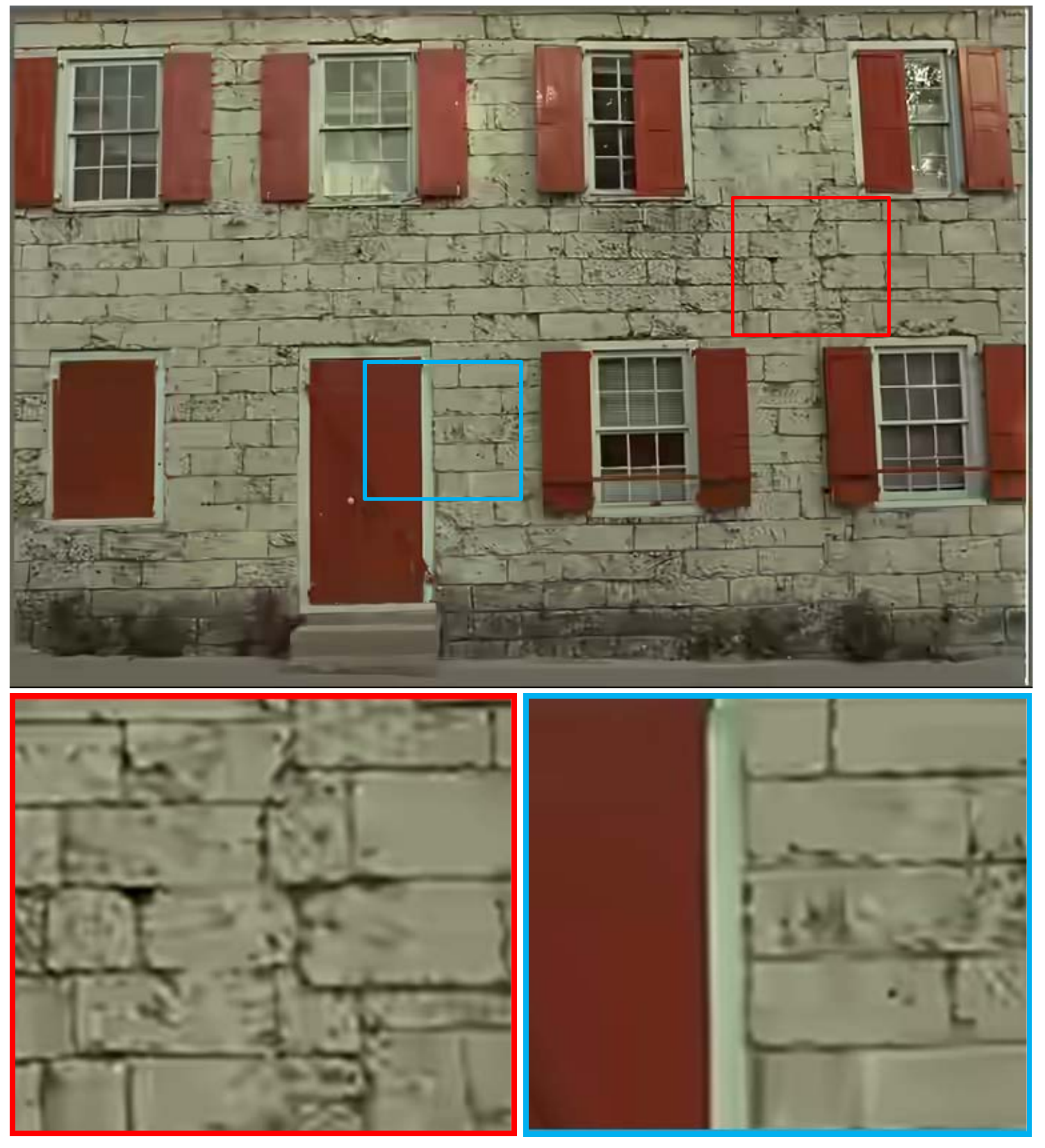}
\end{minipage}
}%
\subfigure[Ours(0.230/26.781/0.943)]{
\begin{minipage}[t]{0.33\linewidth}
\centering
\includegraphics[scale=0.4]{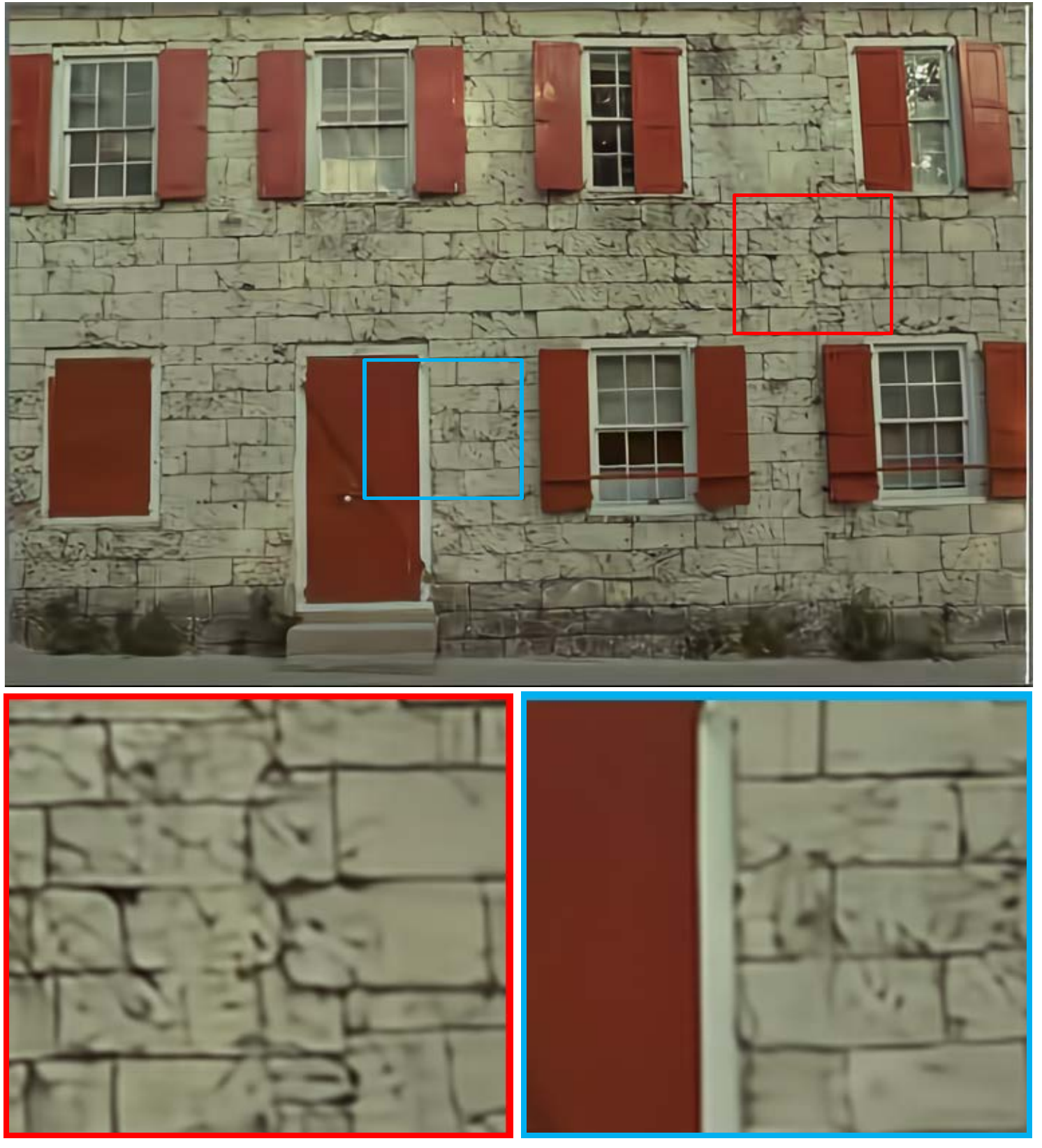}
\end{minipage}
}%
\centering
\caption{Example 1 in the Kodak dataset (bpp, PSNR (dB), MS-SSIM).}
\label{Example1}
\end{figure*}

\begin{figure*}[!tph]
\centering
\subfigure[Original]{
\begin{minipage}[t]{0.33\linewidth}
\centering
\includegraphics[scale=0.6]{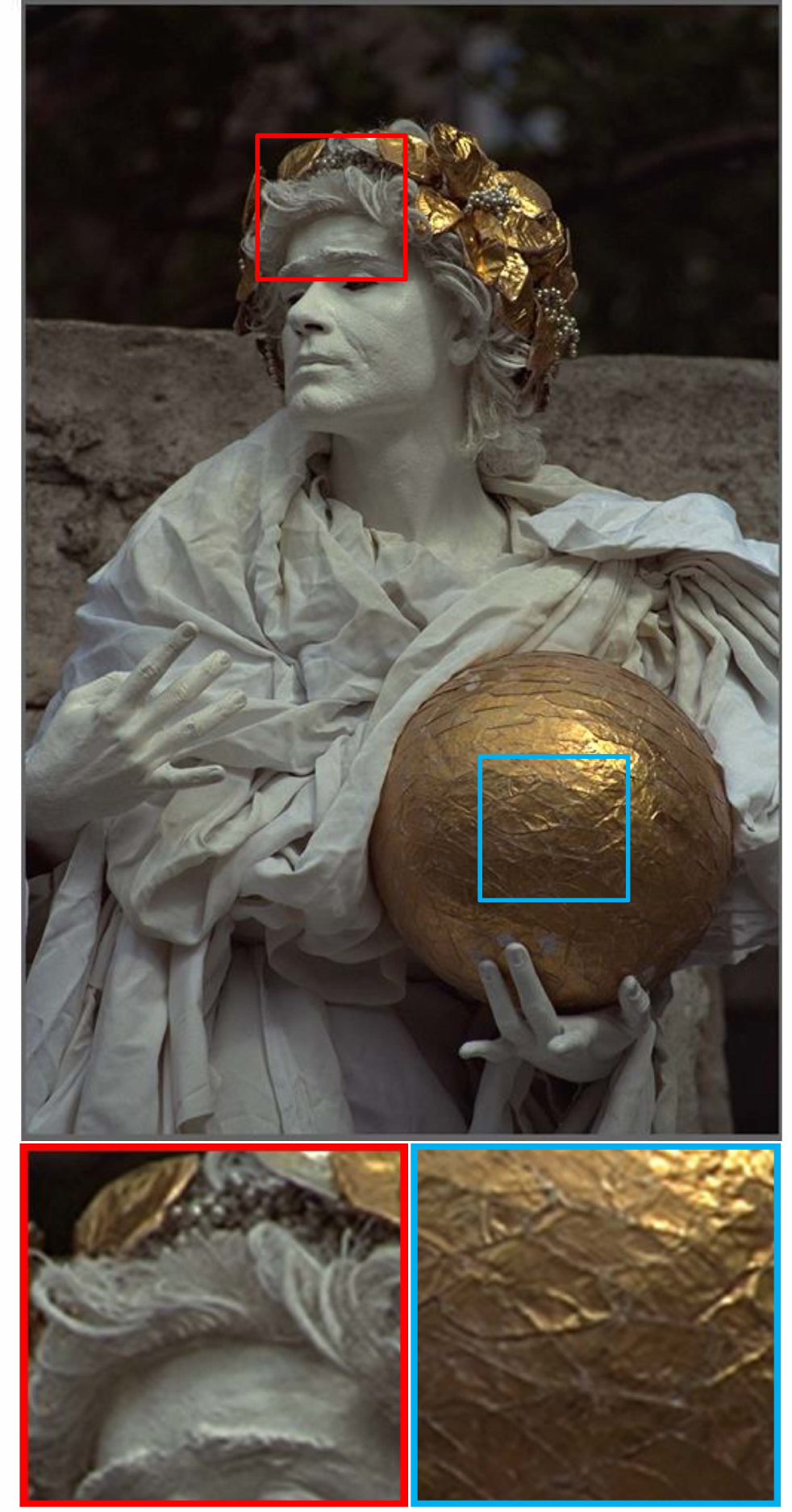}
\end{minipage}
}%
\subfigure[JPEG(0.164/22.98/0.731)]{
\begin{minipage}[t]{0.33\linewidth}
\centering
\includegraphics[scale=0.6]{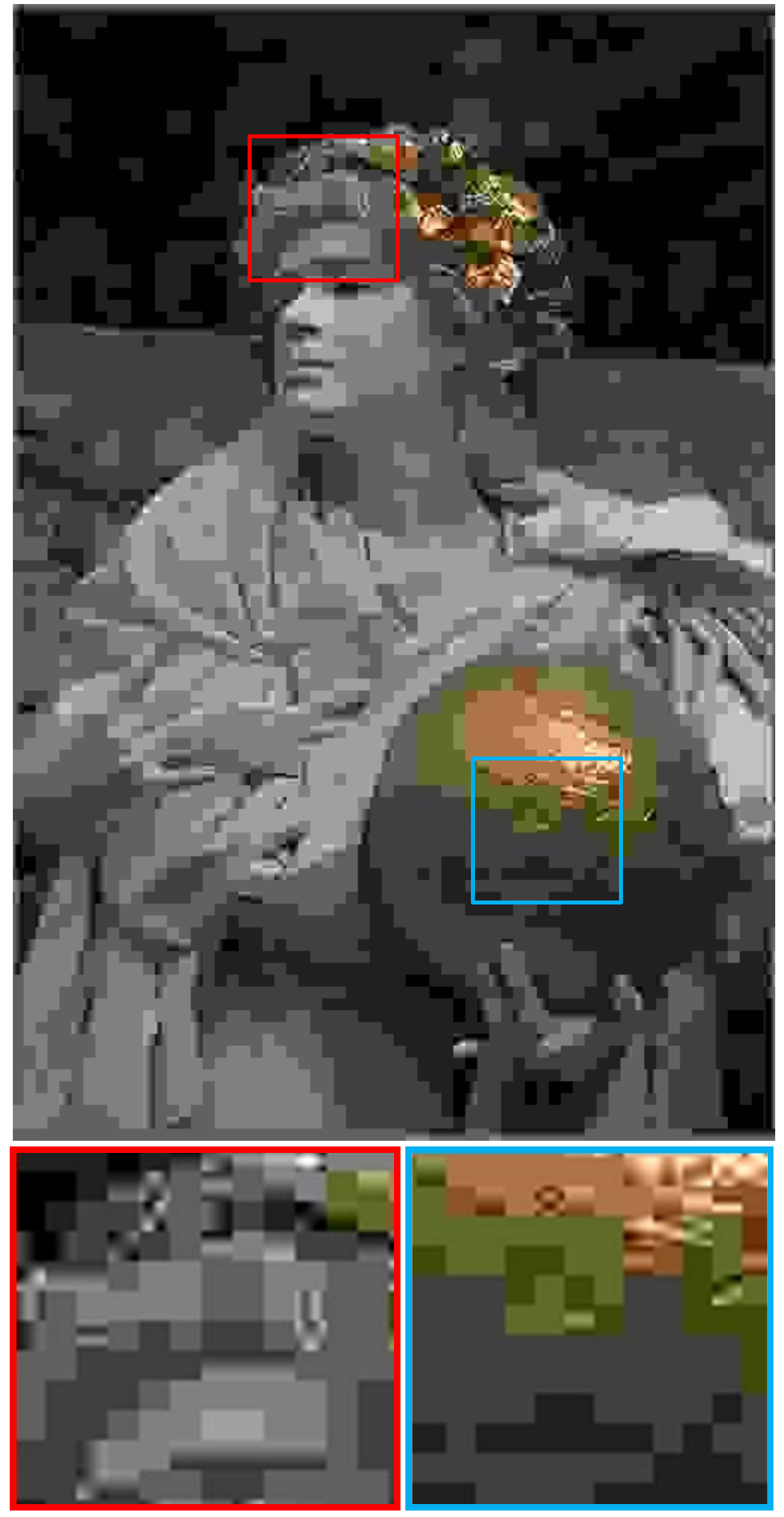}
\end{minipage}
}%
\subfigure[JPEG2000(0.115/28.65/0.924)]{
\begin{minipage}[t]{0.33\linewidth}
\centering
\includegraphics[scale=0.6]{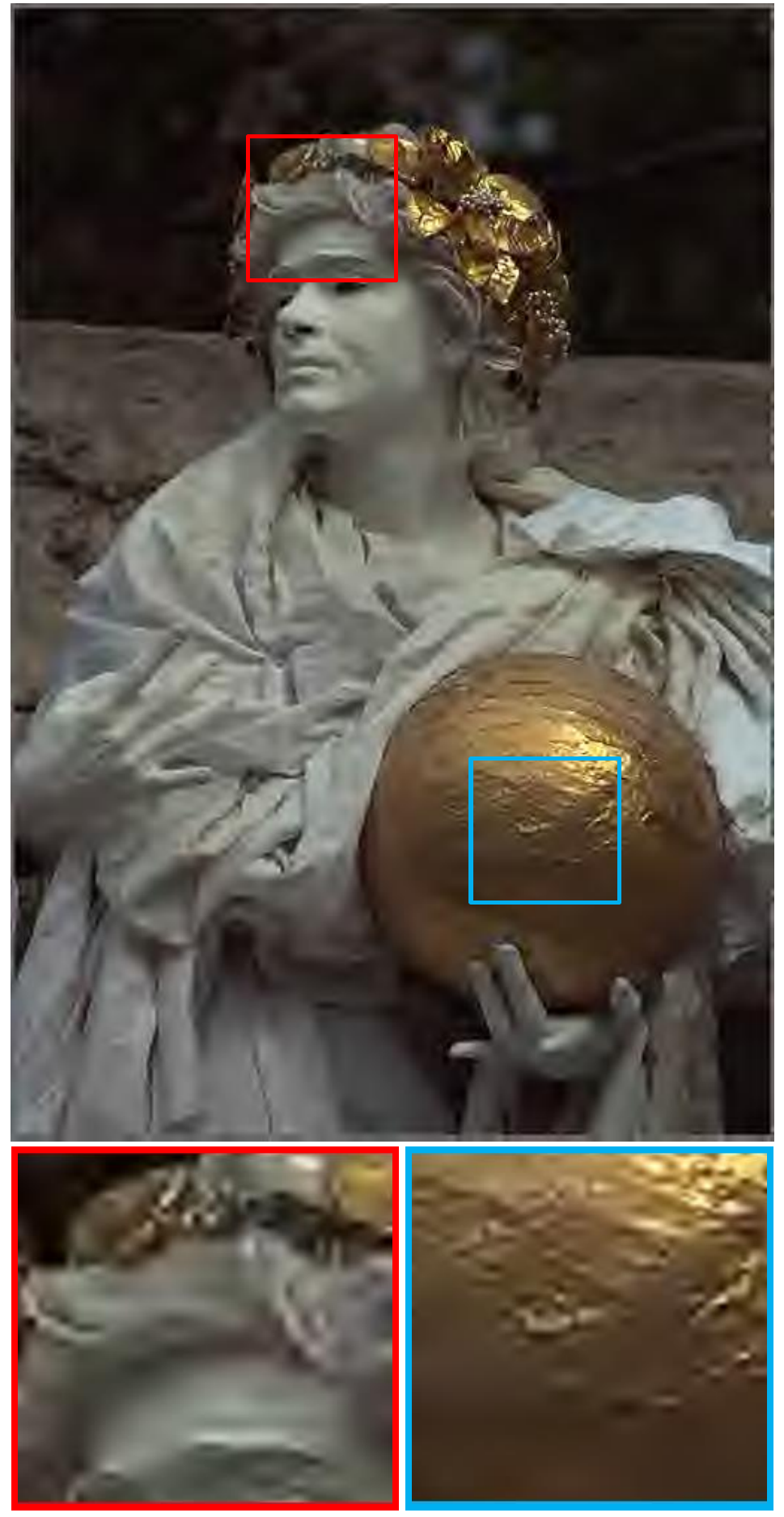}
\end{minipage}
}%

\subfigure[BPG(0.119/29.85/0.938]{
\begin{minipage}[t]{0.33\linewidth}
\centering
\includegraphics[scale=0.6]{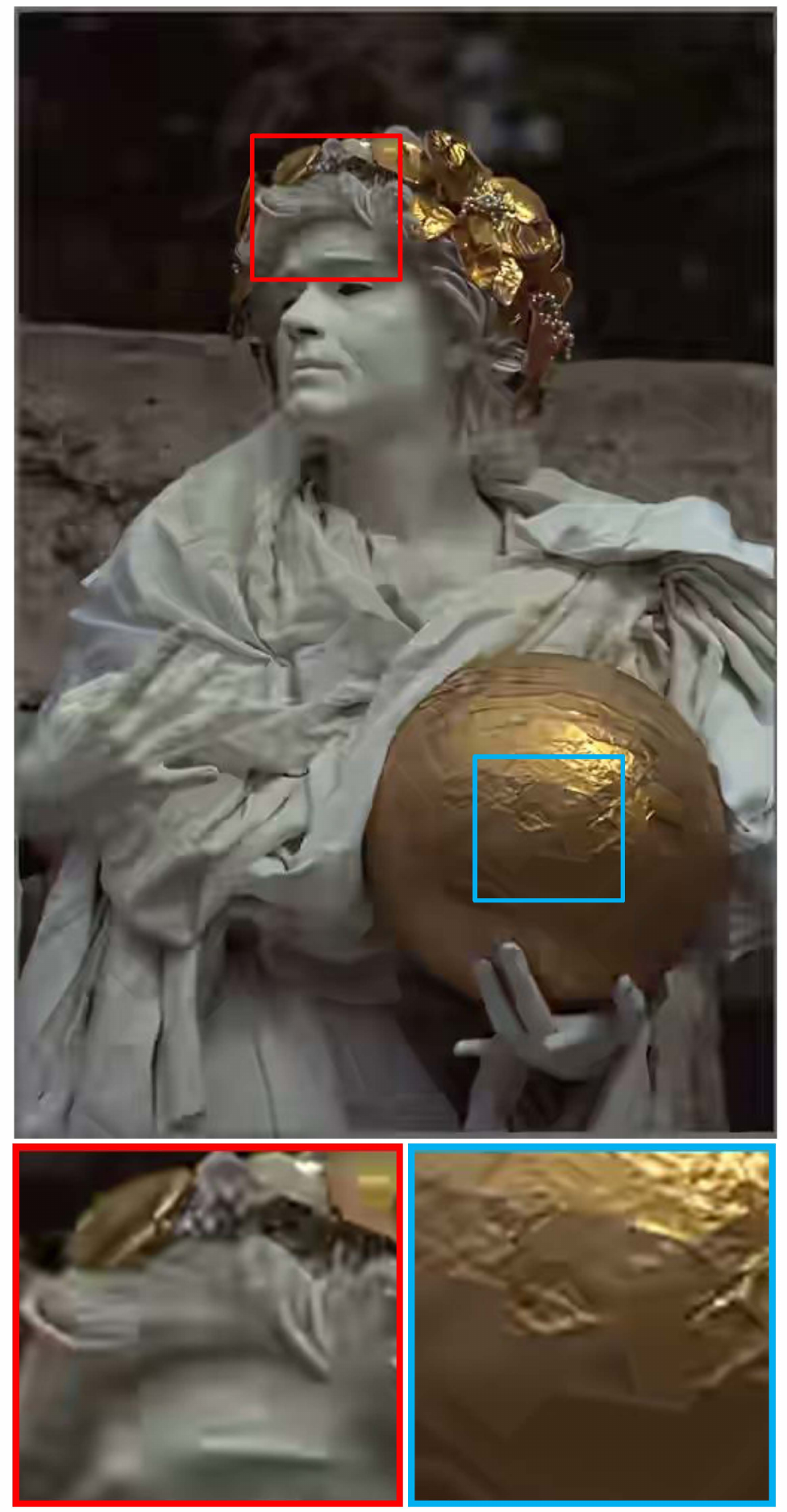}
\end{minipage}
}%
\subfigure[VVC(0.114/30.54/0.943)]{
\begin{minipage}[t]{0.33\linewidth}
\centering
\includegraphics[scale=0.6]{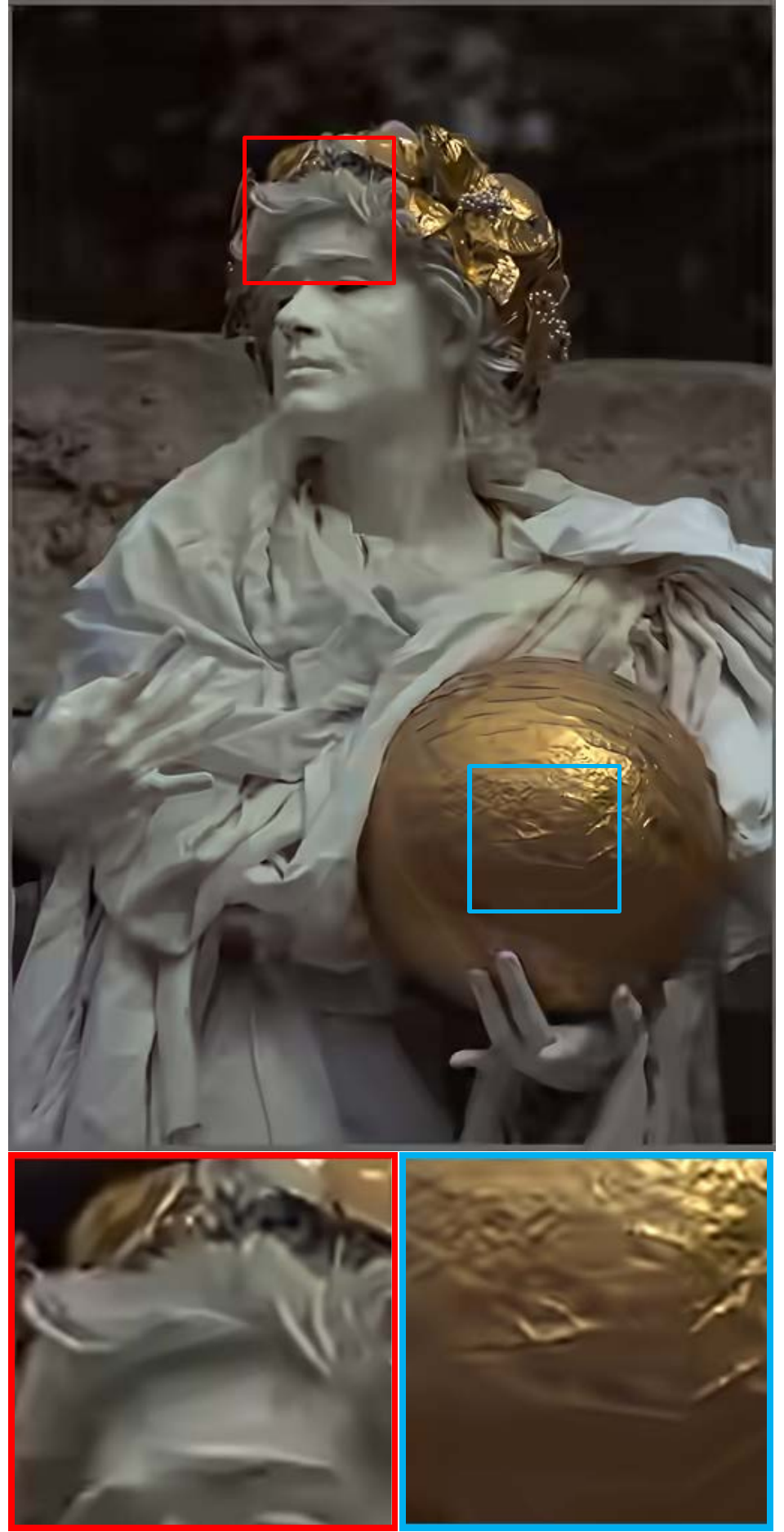}
\end{minipage}
}%
\subfigure[Ours(0.116/30.60/0.952)]{
\begin{minipage}[t]{0.33\linewidth}
\centering
\includegraphics[scale=0.6]{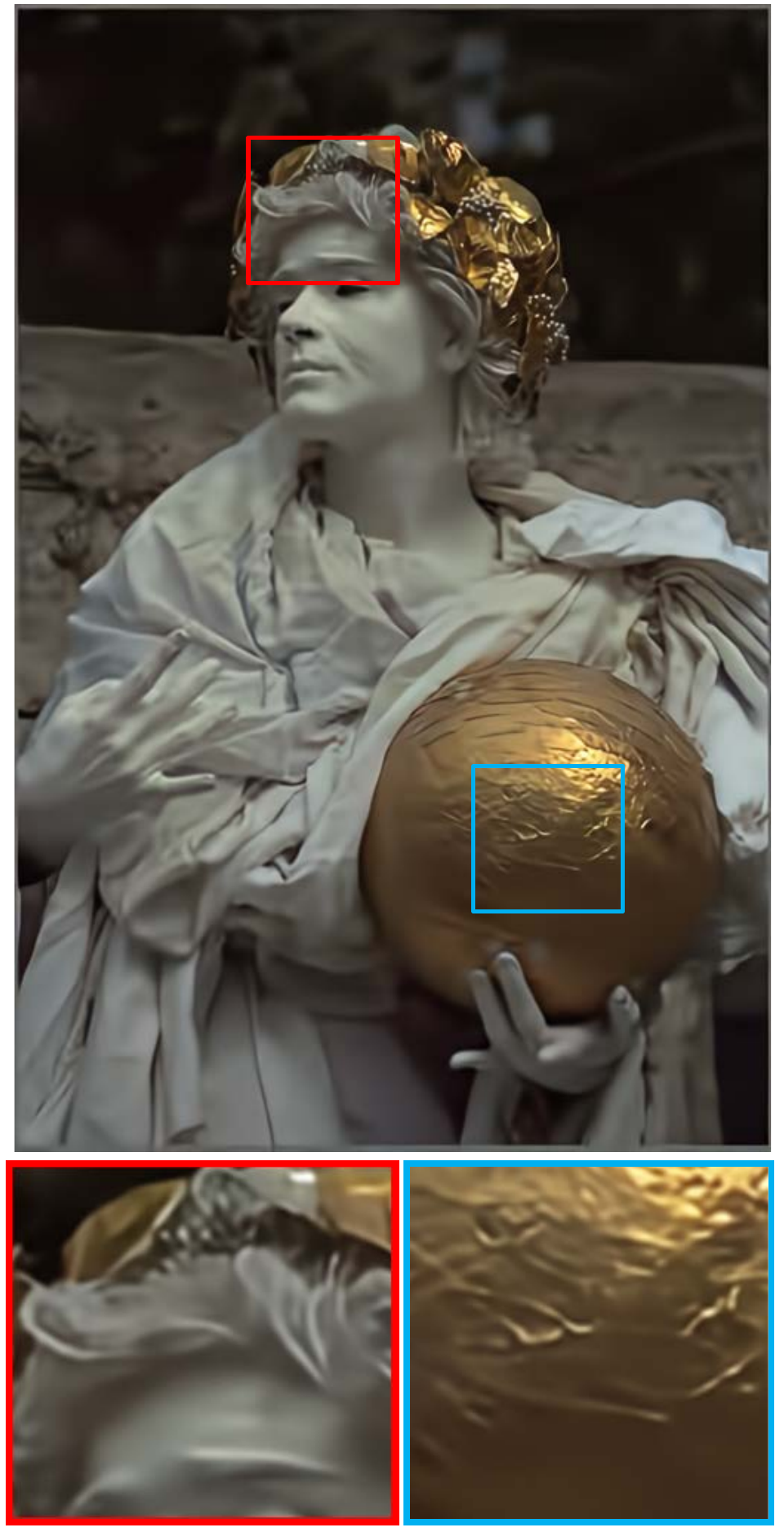}
\end{minipage}
}%
\centering
\caption{Example 2 in the Kodak dataset (bpp, PSNR (dB), MS-SSIM).}
\label{Example2}
\end{figure*}

Fig. \ref{Example1} and Fig. \ref{Example2} is two examples to compare the visual quality of an image coded by our method (optimized for MSE) and the classical image codecs including JPEG, JEPG2000, BPG (4:4:4), and VVC (4:4:4). It is clear that our method preserves more details and achieves the better visual quality than other codecs.

\section{Conclusion}
\label{Conclusion}

 In this paper, our goal aims to balance the model complexity and R-D performance. We propose an efficient asymmetric learned image compression method. We first develop an improved multi-scale residual block for the encoding and decoding networks. Three blocks are used in the encoder. We find that only one block is needed in the decoder and it does not affect the image compression performance. To save  bit rates and reduce the decoding time, we propose an improved importance map to adaptively allocate bits to different regions of the image. Moreover, we apply a 2D post-quantization filter in the latent representation domain to reduce the quantization error, similar to the SAO filter in video coding. Experimental results show that compared to the state-of-the-art method in \cite{GLLMM}, the encoding/decoding complexity of the proposed scheme is reduced by 17 times, and the R-D performance is almost identical. Therefore the proposed scheme achieves the new state of the art in learned image coding when considering both the complexity and performance.

 The results in this paper demonstrate that there are significant amount of redundancy in the existing learned image coding frameworks. The complexity of encoder and decoder have different effects on image compression performance. It is possible to further reduce the complexity without hurting its performance. In particular, the proposed asymmetric paradigm is very attractive for real-time image and video applications, and opens up a new approach to reduce the complexity of the system.
%
%
\ifCLASSOPTIONcaptionsoff
  \newpage
\fi

\bibliographystyle{IEEEtran}
\bibliography{egbib}{}
\end{document}